# Deep Learning in Physical Layer: Review on Data Driven End-to-End Communication Systems and their Enabling Semantic Applications

**NAZMUL ISLAM, GRADUATE STUDENT MEMBER, IEEE, SEOKJOO SHIN, SENIOR MEMBER, IEEE**

Department of Computer Engineering, Chosun University, Gwangju 61452, Republic of Korea

CORRESPONDING AUTHOR: Seokjoo Shin (e-mail: sjshin@chosun.ac.kr).

This work was supported by the National Research Foundation of Korea (NRF) grant funded by the Korea government. (MSIT) (RS-2023-00278294).

**ABSTRACT** Deep learning (DL) has revolutionized wireless communication systems by introducing data-driven end-to-end (E2E) learning, where the physical layer (PHY) is transformed into DL architectures to achieve peak optimization. Leveraging DL for E2E optimization in PHY significantly enhances its adaptability and performance in complex wireless environments, meeting the demands of advanced network systems such as 5G and beyond. Furthermore, this evolution of data-driven PHY optimization has also enabled advanced semantic applications across various modalities, including text, image, audio, video, and multimodal transmissions. These applications elevate communication from bit-level to semantic-level intelligence, making it capable of discerning context and intent. Although the PHY, as a DL architecture, plays a crucial role in enabling semantic communication (SemCom) systems, comprehensive studies that integrate both E2E communication and SemCom systems remain significantly underexplored. This highlights the novelty and potential of these integrative fields, marking them as a promising research domain. Therefore, this article provides a comprehensive review of the emerging field of data-driven PHY for E2E communication systems, emphasizing their role in enabling semantic applications across various modalities. It also identifies key challenges and potential research directions, serving as a crucial guide for future advancements in DL for E2E communication and SemCom systems.

## I. INTRODUCTION

Since the initial conceptualization of neural networks (NNs) in 1943 [1], the field of machine learning (ML) has experienced over seven decades of rigorous research and technological development. The 2010s marked a pivotal era with significant advancements in parallel computing and the emergence of deep learning (DL) paradigms. DL has significantly shifted the paradigm from model-driven to data-driven algorithms. This transition is particularly beneficial in tackling tasks that are challenging to express or solve with traditional mathematical models. The shift is made possible by the availability of extensive, high-dimensional datasets, allowing for more nuanced and sophisticated algorithmic approaches. Continuing the trajectory, various specialized DL architectures have been breaking grounds in numerous domains, such as convolutional neural networks (CNNs) for image processing tasks, reinforcement learning (RL) algorithms for complex decision-making algorithms, generative models for realistic data synthesis, autoencoders (AEs) for dimensionality reduction or anomaly detection, and transformers in natural language processing (NLP) applications. These advancements have set new benchmarks across a wide array of scientific and industrial applications.

In the meantime, new generations of mobile communication systems have been introduced approximately every decade since 1979 [2]. Each new generation of wireless communication systems aims to enhance the accurate transmission of data through wireless channels. These advancements, often characterized by improved reliability and data rates, owe much to the mathematical modeling of communication systems and wireless channels, enabling more precise and efficient data delivery. The mathematical models enabled the design of algorithms using knowledge of signal processing and information theory. As the communication systems became increasingly complex, the tractability of the mathematical models was achieved by dividing the communication system into small blocks, known as processing blocks. Each of these blocks performs a distinct and specific sub-function, such as source coding, modulation, channel estimation, and equalization, as illustrated in Fig. 1(a) [3], [4]. However, such block-based systems often have two major problems, firstly, joint optimization of transmitter and receiver becomes highly intractable for complex and realistic channel models, and secondly, simple channel models fail to account for practical propagation intricacies and hardware impairments. Therefore, individual blocks need to be analyzed and optimized independently. While this approach has led to the optimization of individual blocks based on robust mathematical principles, resulting in stable and prevalent systems in current practical applications, it still falls short of optimal efficiency for the overall communication system. This limitation arises because the system, when divided into separate processing blocks, may not achieve the best possible performance as a unified whole. Additionally, the signaling between the transmitter and receiver introduces extra overhead, consequently lowering the overall throughput of the communication systems.

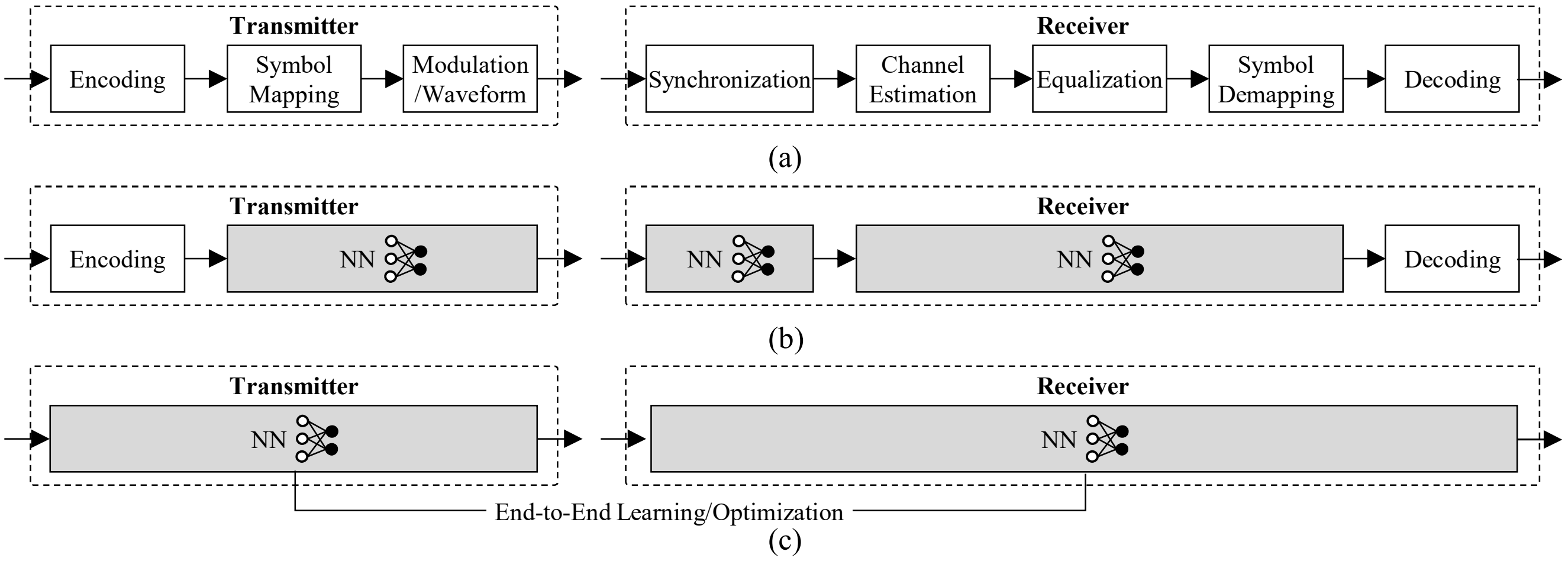

**FIGURE 1. PHY in communication system. (a) A traditional block-based system, (b) DL-based block optimization where one or multiple blocks are replaced with NN and (c) DL-based E2E optimization of transceiver.**

The study of DL in the physical layer (PHY) of communication systems dates to the 1990s. With the recent advances in artificial intelligence (AI), interest in this area has experienced a resurgence, indicating a renewed focus on integrating DL into the framework of communication systems [5]. Various studies have investigated model-driven DL-based communication systems, where one or multiple blocks are replaced with deep neural networks (DNNs), as shown in Fig. 1(b). This requires the next generation of communication systems to be designed in a way that allows for training and testing of DNNs in each block. Although this method presents a novel approach, it optimizes each block separately rather than focusing on maximizing the overall system performance. This independent optimization may not yield the most efficient results for the entire communication system. On the other hand, an end-to-end (E2E) communication system offers a paradigm shift from the traditional communication system design without relying on separate communication blocks. In this setup, the PHY is realized through DNN coupled with E2E learning, striving to achieve highly accurate data reproduction, as depicted in Fig. 1(c). E2E systems promise learning to communicate across real-world environments using data-driven optimization without the prerequisite of mathematical modeling and intricate analysis. Since first introduced in [6], the E2E communication system has been extended to various fields of communication, including optical wireless [7], optical fiber [8], and practical systems [9]. Furthermore, E2E systems have led to integration of fields in signal processing, information theory, and DL in communication systems. For instance, the transceiver can be conceptualized as an AE where the task of estimating transmitted bits is approached as a binary classification problem. Moreover, the achievable rate of communication system can be quantified using cross entropy, a well-known metric in both DL and information theory. These advancements have led the E2E communication system to revolutionize the PHY, enabling both the transmission and reception processes through DL, thereby enhancing the overall efficiency and reliability of the system.

Owing to its data-driven approach, E2E communication has laid the foundation for semantic communication (SemCom) systems, which are adapted for semantic applications such as image transmission, speech transmission, text transmission, video transmission, and multimodal transmission. Unlike traditional communication systems which focus on bit-level or symbol-level transmission efficiency, SemCom focuses on meaning of the information based on the context of communication. As noted by Shannon and Weaver [3], communication can be divided into three fundamental levels: (1) Technical level – how accurate are the symbol transmissions? (2) Semantic level – how exactly does the symbol transmission convey the desired meaning? and (3) Effectiveness level – how effective is the conveyed meaning to achieve the desired outcome? The study of semantics, a field that has gained much attention for centuries across various disciplines, essentially focuses on the interpretation of 'meaning'. However, it is important to note that the way meaning is understood and analyzed can significantly differ based on the specific domain of study. In communication systems, SemCom aims to optimize data transmission at the semantic level. By extracting relevant semantic information and minimizing data, this approach allows for higher data compression, lower signal-to-noise ratio (SNR), and improved performance within the same bandwidth.

Emerging AI-based applications such as autonomous driving and smart cities highlight the need for intelligent, data-aware network systems that understand the relevance, urgency, and semantics of transmitted data in relation to specific tasks. These applications depend on massive datasets that are used for training extensive models for various functions. Transmitting these datasets can produce substantial traffic, potentially overwhelming network capacity, while also demanding extremely low E2E latency. SemCom addresses these challenges of optimizing network capacity by eliminating the need for bit-by-bit reconstruction of information at the receiving end. It seeks to minimize the number of transmitted symbols by only sending bits associated with the meaning of the data and incorporating the context of the communication. Additionally, SemCom can achieve greater data compression without compromising its meaning, leading to a considerable reduction in data transmission. Therefore, when compared to traditional communication systems, SemCom can operate with reduced SNR or bandwidth, and achieve better transmission performance under identical conditions. In recent years, the advances in AI, pervasive networks, and communication systems have intensified the interest in semantic and goal-oriented communications, particularly with the emergence of next-generation systems like sixth-generation (6G) wireless communication systems.

## II. RELATED WORK AND CONTRIBUTION

With the advancements of semantic communication (SemCom) and deep learning (DL)-based techniques in modern communication systems, there are several surveys discussing machine learning (ML) in communication systems and SemCom. For instance, article [10] delves into the theory, framework, and system design of SemCom, emphasizing performance metrics distinct from traditional communication systems. In [11], the focus is on DL-based methods in the PHY for prospective 6G applications, highlighting key PHY concepts like massive multiple-input multiple-output (MIMO) systems, multi-carrier waveform design, reconfigurable intelligent surface-empowered communications, and security in the physical layer (PHY). The tutorial [12] provides an overview of the progress made to date in SemCom, beginning with its initial adaptations and spanning semantic-driven and task-focused communications. It delves into the underlying principles, methodologies, and possible applications. In article [13], the authors emphasize role of SemCom in 6G, detailing its types such as semantic-oriented, goal-oriented, and semantic-aware communication, and system design aspects in terms of extracting semantic information, its transmission, and related metrics. The paper also showcases the potential applications of SemCom in 6G and provides insights into its possible impact on future network architectures. Article [14] provides a comprehensive overview of semantic transmissions, delving into their history, significance, and related research areas. It categorizes critical techniques, highlights their evolution in modern communications, and explores their connection to broader communication contexts. Additionally, the article emphasizes advanced methods to improve semantic accuracy and scalability in practical applications. In [15], the evolution of SemCom is outlined, emphasizing its features and primary technologies. Solutions concerning efficiency and adaptability are highlighted. The application of SemCom in areas like unmanned aerial vehicle (UAV) communication, remote sensing, transportation, and healthcare is also discussed. In [16], the authors primarily aim to highlight recent developments in DL-based algorithm techniques within the PHY of wireless communication. It also discusses their prospective applications in upcoming communication systems. In addition, several overviews, such as [17]–[21], provide diverse perspectives on SemCom design, covering topics ranging from ML technologies in

**TABLE 1. Summery and comparison of existing literature review**

| Ref. | Year | Brief Description | PHY as E2E System | | SemCom Applications | | | | | |
|---|---|---|---|---|---|---|---|---|---|---|
| | | | E2E System Description | E2E In Communication | SemCom Description | Text | Image | Audio | Video | Multi-modal |
| [10] | 2022 | Theory, framework, and system design of SemCom | ○ | ○ | ● | ● | ● | ● | ● | ● |
| [11] | 2022 | DL methods in the PHY for prospective 6G application | ● | ● | ○ | ○ | ○ | ○ | ○ | ○ |
| [17] | 2022 | Overview on machine learning technology in wireless communication systems | ● | ○ | ● | ○ | ○ | ○ | ○ | ○ |
| [12] | 2023 | Overview of SemCom with initial adaptations, and semantic-driven and task-focused communications | ○ | ○ | ● | ● | ● | ○ | ● | ○ |
| [13] | 2023 | Type of SemCom in 6G, such as semantic-oriented, goal-oriented, and semantic-aware communication | ○ | ○ | ● | ● | ● | ● | ○ | ● |
| [14] | 2023 | Comprehensive overview of semantic transmissions, its history, significance, and related research areas | ○ | ○ | ● | ● | ● | ● | ● | ○ |
| [15] | 2023 | The evolution of SemCom is outlined, emphasizing its features and primary technologies | ○ | ○ | ● | ● | ● | ○ | ○ | ○ |
| [16] | 2023 | Recent developments in DL-based algorithm techniques within the physical layer of wireless communication | ● | ● | ● | ○ | ○ | ○ | ○ | ○ |
| Current Study | | Comprehensive review of E2E communication systems and its enabling semantic applications | ● | ● | ● | ● | ● | ● | ● | ● |

Notation indications: ● covered in the study; ○ not covered in the study.

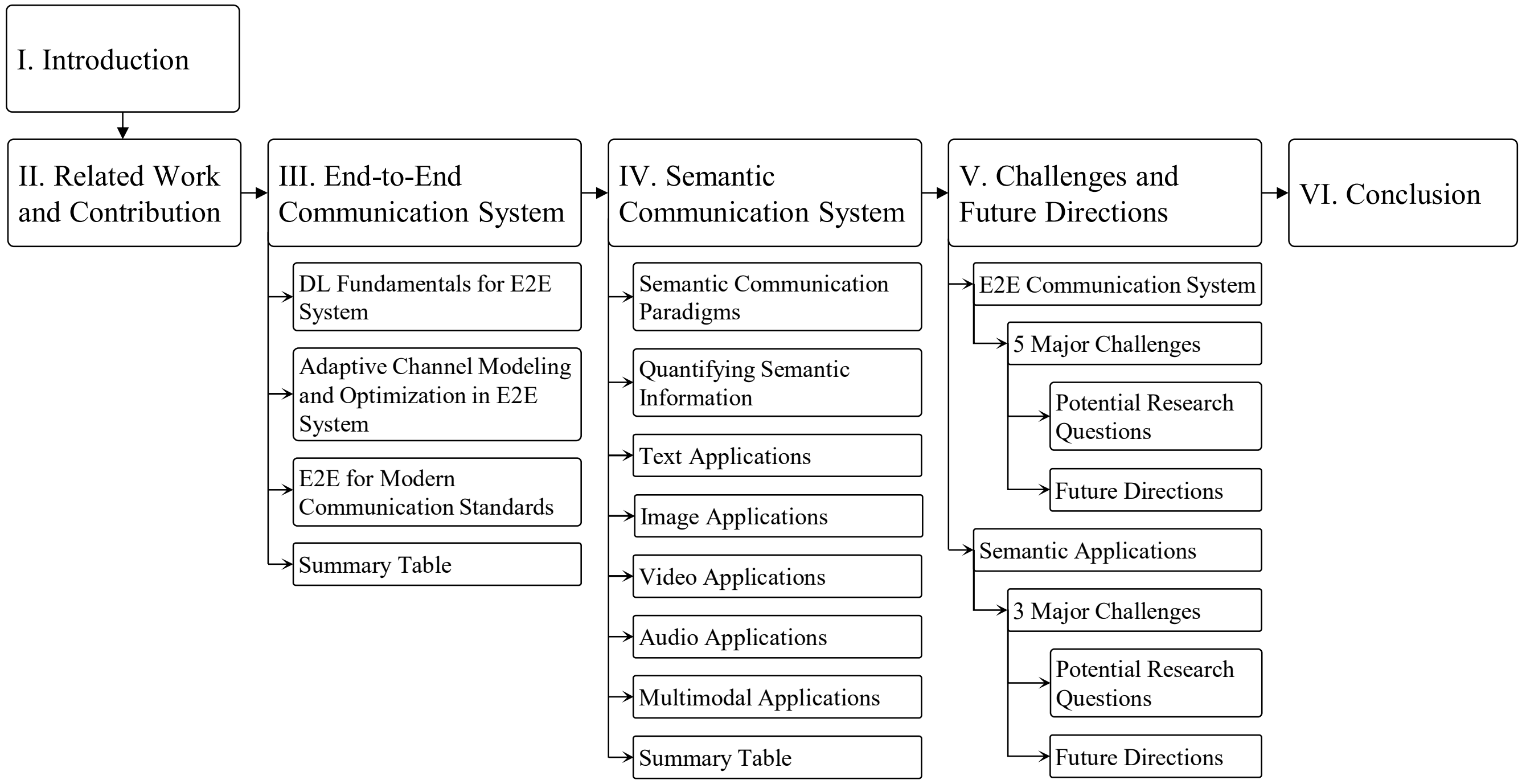

**FIGURE 2. Structural organization of the article.**

wireless communication systems [17], to federated learning (FL)-enabled SemCom networks [18], DL techniques for future 6G communications in the PHY [19], DL-enabled end-to-end (E2E) semantic networks [20], and SemCom-empowered ubiquitous-X 6G frameworks [21]. A summary of existing literature reviews is shown in Table I.

While DL-based PHY and SemCom technologies have garnered significant attention in recent years, a combined review of technologies, solutions, and applications for both areas remains underexplored. The data-driven approach in E2E systems has paved the way for SemCom, and has been adapted for various semantic applications, such as image, audio, video, text, and multimodal transmission. Previous studies focus on either SemCom or DL-based PHY exclusively, without providing a holistic review that bridges the two domains. This article offers a comprehensive review of various E2E technologies, followed by an overview of SemCom and its applications in image, speech, text, video, and multimodal transmission. The main contributions are:

- *DL-based PHY for E2E Communication:* The article first reviews the DL techniques in PHY to develop data-driven E2E communication systems. Furthermore, it discusses how DL-based E2E learning, and optimization address the drawbacks of existing multicarrier waveform systems such as orthogonal frequency division multiplexing (OFDM). The proposed system models are then summarized with key contributions and limitations.
- *Semantic Applications enabled by E2E Learning:* The article then provides a comprehensive review of semantic applications enabled by E2E learning across various modalities (image, text, audio, video, and multimodal transmissions). Varying communication environments are highlighted with key ideas and limitations of the systems.
- *Challenges and Future Directions in E2E and SemCom:* Finally, the article addresses the challenges in data-driven E2E systems and SemCom. It identifies the gaps in current research and provides insights into the potential future trends toward the promising 6G.

The remainder of the article is organized as follows. Section III reviews DL-based PHY for E2E communication. Section IV surveys SemCom applications, comparing and discussing various DL-based approaches. Challenges and future directions are discussed in Section V. Finally, Section VI. concludes the paper. Fig. 2 provides the organizational structure of the article with main sections and covered topics.

## III. END-TO-END (E2E) COMMUNICATION SYSTEM

Traditional communication systems face several limitations, including inaccuracies in channel state information (CSI) in complex transmission environments, exponentially increasing complexity when processing large volumes of data, and suboptimal performance in conventional block-based systems. With the advancements in artificial intelligence (AI), many researchers have leveraged machine learning (ML), particularly deep learning (DL) techniques, to develop advanced communication systems that can outperform traditional systems [9]. Unlike traditional communication systems, which are composed of multiple discrete blocks like source/channel encoding, modulation, channel assessment, and equalization, DL-driven systems integrate these functions. This integration allows the DL-based system to simultaneously optimize both the transmitter and receiver, enhancing E2E performance. Specifically, it implements the system as a unified deep neural network (DNN) as shown in Fig. 3, improving overall communication efficacy [6], [22]. This implementation offers significant benefits. Firstly, it streamlines end-to-end (E2E) performance by utilizing DNNs, which replace the block based design typical of traditional systems. Secondly, it optimizes the system for any channel model, even those without a precise mathematical representation, which includes complex and non-linear transmission scenarios. Lastly, it harnesses parallel neural network (NN) processing on concurrent architectures, enabling faster processing times compared to conventional algorithms, with the added advantage of functioning with low-precision data types [23]. To develop an E2E system, the communication system is initially configured as an autoencoder (AE), which is trained with a pre-compiled offline dataset. The AE is subsequently integrated into operational online systems, undertaking the task of E2E reconstruction. This process simultaneously optimizes the transmitter and receiver and learns the signal encoding, thereby enhancing system efficiency and performance [6].

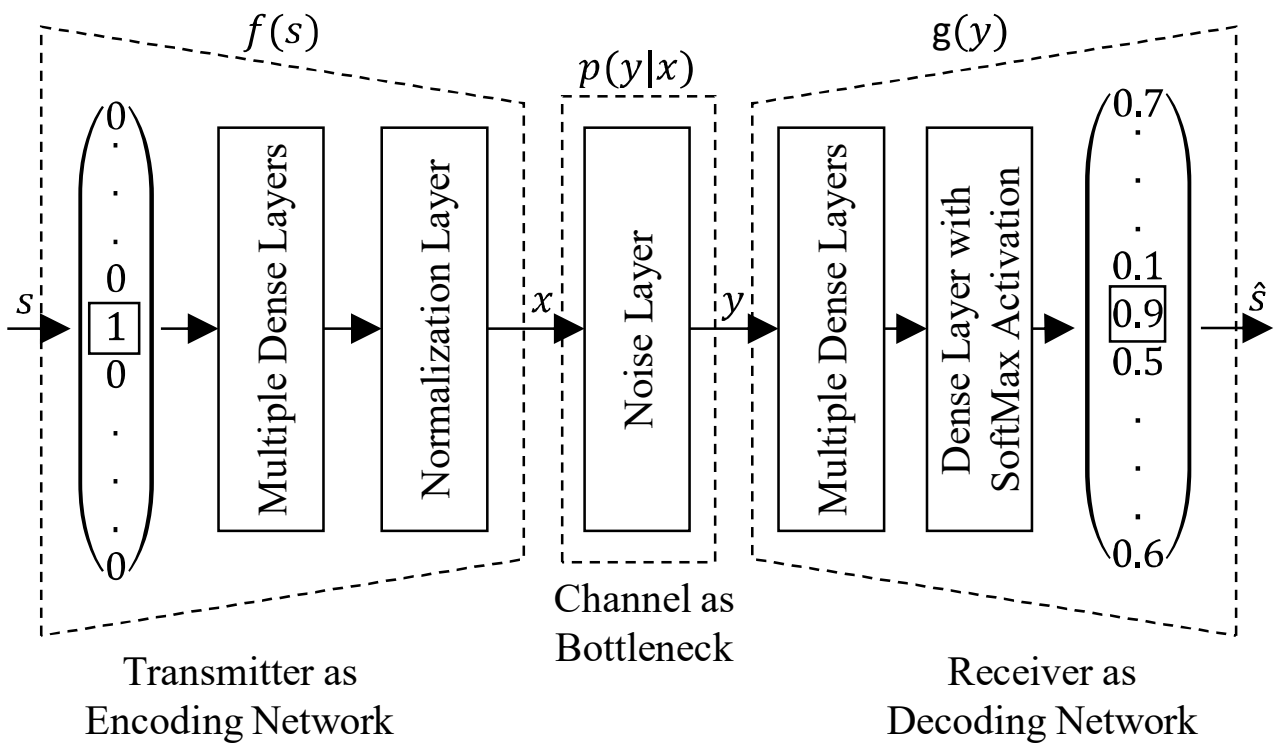

**FIGURE 3. AE based E2E communication system. Input $s$ is encoded using one-hot vector, and output is probability distribution with most probable message as output $\hat{s}$ [6].**

### A. DL FUNDAMENTALS FOR E2E SYSTEM

A multilayer perceptron (or feed forward NN) consisting of $L$ layers represent mapping $f(r_0;\theta)\colon \mathbb{R}^{N_0} \to \mathbb{R}^{N_L}$ of input $r_0 \in \mathbb{R}^{N_0}$ onto an output $r_L \in \mathbb{R}^{N_L}$ using $L$ iterations:

$$r_l = f_l(r_{l-1};\theta_l), \qquad l = 1, \dots, L \tag{1}$$

where the $l^{th}$ layer performs the mapping $f_l(r_{l-1};\theta_l)\colon \mathbb{R}^{N_{l-1}} \to \mathbb{R}^{N_l}$. Along with the output vector $r_{l-1}$ from previous layer, the mapping also depends on the $\theta_l$ set of parameters. This mapping can be stochastic with $f_l$ as a random variable function. $\theta = \{\theta_1, \dots, \theta_L\}$ denotes all the parameter sets of the network. The $l^{th}$ layer is the *fully-connected* (*dense*) layer if $f_l(r_{l-1};\theta_l)$ is:

$$f_l(r_{l-1};\theta_l) = \sigma(W_l r_{l-1} + b_l). \tag{2}$$

The parameter set for this dense layer is $\theta_l = \{W_l, b_l\}$. $W_l \in \mathbb{R}^{N_l \times N_{l-1}}$, $b_l \in \mathbb{R}^{N_l}$ and $\sigma(.)$ is the *activation function*. Several other layer types with their mapping function are:

$$f_l(r_{l-1};\theta_l) = r_{l-1} + n, n \sim \mathcal{N}\left(0, \beta I_{N_{l-1}}\right), \tag{3}$$

$$f_l(r_{l-1};\theta_l) = d \odot r_{l-1}, d_i \sim Bern(\alpha), \tag{4}$$

**TABLE 2. Activation functions**

| Name | $[\boldsymbol{\sigma}(\boldsymbol{u})]_i$ | Rage |
|---|---|---|
| Linear | $u_i$ | $(-\propto, \propto)$ |
| RELU | $max(0, u_i)$ | $(0, \propto)$ |
| Tanh | $tanh(u_i)$ | $(-1,1)$ |
| Sigmoid | $\frac{1}{1+e^{-u_i}}$ | $(0,1)$ |
| SoftMax | $\frac{e^{u_i}}{\sum_j e^{u_j}}$ | $(0,1)$ |

$$f_l(r_{l-1};\theta_l) = \frac{\sqrt{N_{l-1} r_{l-1}}}{||r_{l-1}||_2}. \tag{5}$$

(3) is for noise, (4) is for dropout, and (5) is for normalization.

Each time the layers with stochastic mappings are called, it generates a new random mapping. For instance, noise layers add Gaussian vectors (with zero mean and $\beta I_{N_{l-1}}$ covariance matrix) to the input, and each time they are called for the same input, they generate a different output. The activation function introduces non-linearity, crucial for expressive capabilities of the NN. Without this non-linearity, simply stacking multiple layers would not enhance the network functionality. Typically, the activation function is applied individually to each component of the input vector, exemplified as $[\sigma(u)]_i = \sigma(u_i)$, where σ represents the activation function applied to each input unit $u_i$. A few common activation functions with their range are given in Table II. NNs are usually trained using labelled data, which is a set of input-output vectors $\left(r_{0,i}, r^*_{L,i}\right), i = 1, \dots, S$, where $r^*_{L,i}$ represents the desired output corresponding to the input $r_{0,i}$. The primary objective of training is to minimize the loss function with respect to the parameters denoted by $\theta$:

$$L(\theta) = \frac{1}{S}\sum_{i=1}^{S} l(r^*_{L,i}, r_{L,i}), \tag{6}$$

where $r_{0,i}$ is the input with the corresponding output $r_{L,i}$, and $l(u,v)\colon \mathbb{R}^{N_L} \times \mathbb{R}^{N_L} \to \mathbb{R}$ is the loss function. Most common loss functions used in E2E are mean squared error (MSE), $l(u,v) = ||u-v||_2^2$ and categorical cross-entropy, $l(u,v) = \sum_j u_j \log(v_j)$. Different regularization (such as L1 and L2) can be used to achieve greater performance with less or distributed values. One of the most common algorithms to find optimal parameter sets $\theta$ is stochastic gradient decent (SGD), which initially starts with random values $\theta = \theta_0$ and updates the sets $\theta$ iteratively as:

$$\theta_{t+1} = \theta_t - \eta \nabla \tilde{L}(\theta_t), \tag{7}$$

where $\eta$ (greater than 0) is the learning rate and $\tilde{L}(\theta_t)$ is the loss function approximation, calculated from random *mini-batch* from the training example $S_t \subset \{1,2,\dots,S\}$ at the iteration:

$$\tilde{L}(\theta_t) = \frac{1}{S_t}\sum_{i \in S_t} l\left(r^*_{L,i}, r_{L,i}\right), \tag{8}$$

$S_t$ is intentionally smaller than $S$ to simplify the computation of the gradient and to decrease the variance in the weight

update, as cited in [6]. There are various types of SGD algorithms which can improve convergence by dynamically adapting to the learning rate. Back propagation algorithms can efficiently compute variants in Eq. (8).

### B. ADAPTIVE CHANNEL MODELING AND OPTIMIZATION IN E2E SYSTEM

A significant drawback of the E2E system is the need for channel state information, or the channel transfer function, for optimizing the transmitter. The authors in [24] categorize channels in AE-based E2E systems as either model-assumed, where statistical models account for channel impairments with non-differentiable transfer functions, or model-free, where no specific channel model is presumed and the receiver is fine-tuned based on actual channel conditions. DNN weights are refined using SGD, which allows error gradients to flow from the output (final) back to the input (initial) layer, streamlining the learning process. Moreover, accurately predicting CSI in real-world communication environments is often unfeasible due to unpredictable noise and fluctuations, lacking a clear analytical model. Various solutions have been proposed by researchers to overcome this issue, such as the simultaneous perturbation stochastic optimization technique presented in [25], which approximates channel gradients and makes the E2E model trainable using standard back propagation. Additional methods, referenced in [26], [27], employ a strategy of alternating between reinforcement learning (RL)-based transmitter training and supervised receiver training, as depicted in Fig. 4. This technique has been proven to deliver performance on par with E2E supervised DNN networks, even with established differentiable channel models, which were substantiated by tests using software-defined radio (SDR). However, the effectiveness of this method is dependent on having a perfect feedback link. The study in [28] introduced a robust feedback system, effectively reducing the dependency on an ideal feedback link. This system closely matches the performance of an E2E alternating network even in environments with noisy feedback. Nonetheless, the effectiveness of the system partly depends on having some preliminary knowledge about the channel characteristics.

Authors in [29], [30] have developed channel-agnostic E2E network, utilizing a conditional generative adversarial network (GAN) to accurately model the conditional distribution of various channels. This approach leverages pilot signals in dynamic channels to refine and generate more accurate models for current channel conditions. Despite its success in generating practical channel models and conditions, the system model in [29] encountered a dimensionality issue, where the codeword size expanded exponentially with input dimension length, rendering it practical only for smaller blocks. This was overcome in [30] using convolution layers which solved the dimensionality problem in large blocks. Additionally, GAN-based E2E networks encountered vanishing gradient and overfitting problems, which were addressed in [31], [32] using residual-aided GAN. Simulation results indicated that approaches [31], [32] surpassed the performance of the alternating training methods [26], [27] as well as the GAN-based E2E systems [29], [30] in various scenarios. Moreover, the authors in [33], [34] introduced a stochastic convolutional layer to represent the channel model effectively. Simulations conducted in both frequency-selective and flat-fading multiple-input multiple-output (MIMO) channels revealed that this proposed model excels in outperforming least square (LS) and minimum mean squared error (MMSE) channel estimations. However, it does not reach the performance level of the GAN-based E2E network. In [35], authors introduced an E2E communication system featuring a DNN-based channel module, utilizing one-dimensional convolutional neural network (CNN) modules in both the transmitter and receiver to process finite-length sequences. This study effectively addresses the complexity found in conditional GAN systems by adopting a simpler network architecture, significantly enhancing stability in training, and accelerating the rate of convergence.

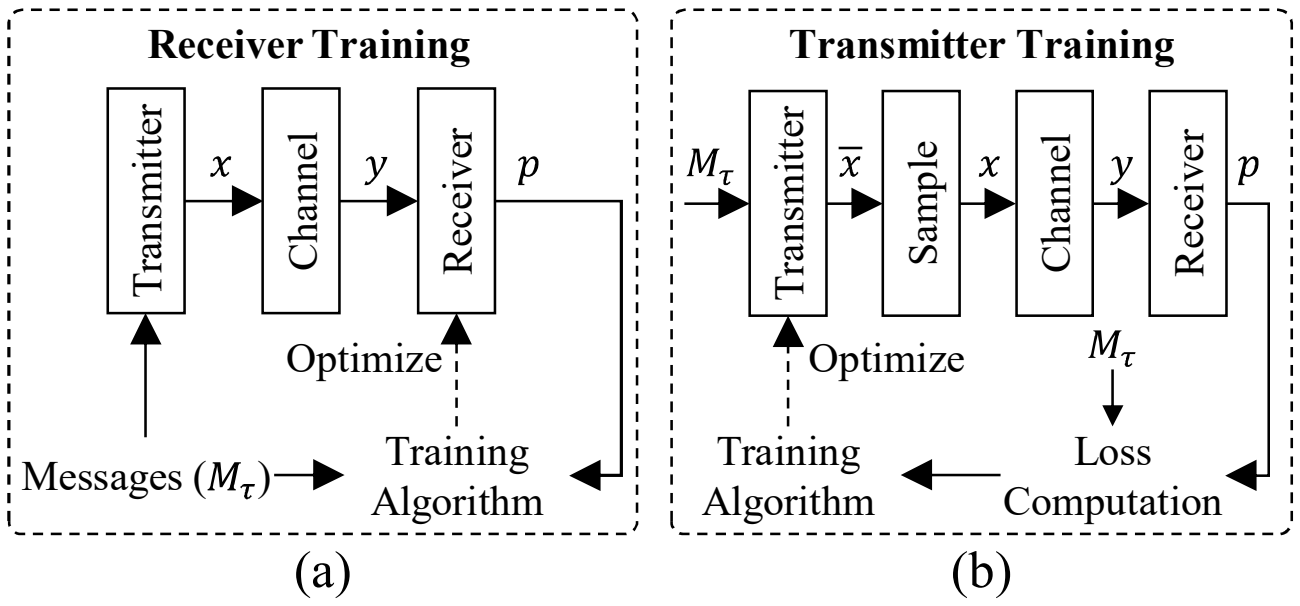

**FIGURE 4. Iteration of (a) receiver and (b) transmitter training of two phases. Where, $p$ is the probability vector of messages, and, $x$ and $y$ are the channe input and output, respectively [27].**

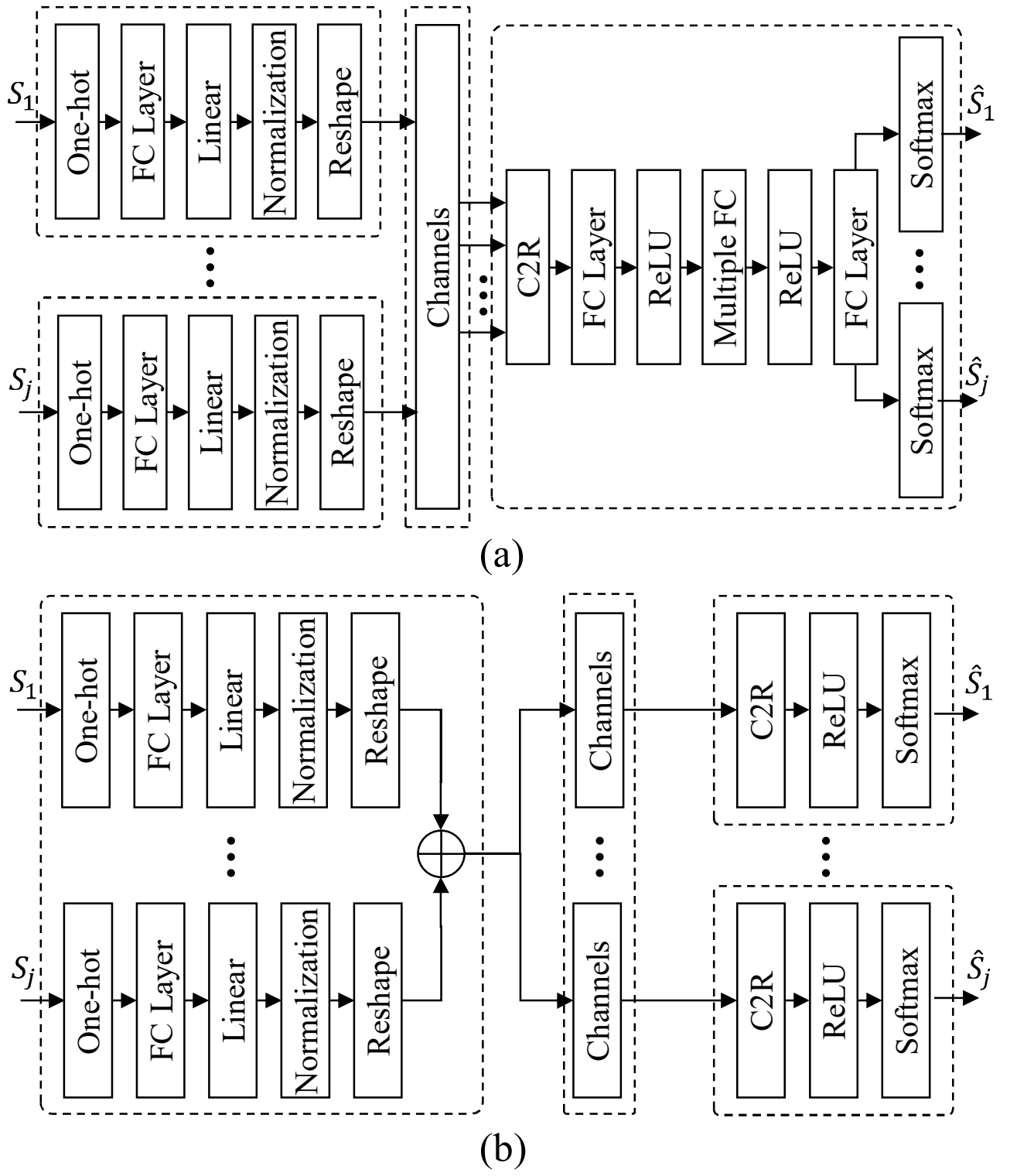

**FIGURE 5. Network structure of (a) uplink and (b) downlink MU-MC-AE. Where, $S$ is the input one-hot encoded vector, and $\hat{S}$ is the output decoded vector of user 1 to $j$ [37].**

### C. E2E SYSTEMS FOR MODERN COMMUNICATION STANDARDS

DL-based methods can be applied to both the transmitter and receiver ends, enabling the optimization of the entire multicarrier waveform systems, such as orthogonal frequency division multiplexing (OFDM) transceiver in an E2E manner. The study in [36] presents an integrated E2E approach in an OFDM system, aiming to jointly optimize both modulation and demodulation operations. Furthermore, research in [37] introduces a DL-based multi-carrier (MC) system using AE, termed MC-AE. This system is designed to optimize modulation and demodulation simultaneously, enhancing both coding efficiency and diversity gains. The architecture of this system, accommodating both uplink and downlink multi-use (MU) and MC transmissions, is depicted in Fig. 5. The performance of communication systems, such as intelligent MIMO transmitters, is significantly impacted by constellation design for modulation. Traditional methods like quadrature amplitude modulation (QAM) and phase shift keying (PSK) modulation schemes are being replaced by complex-shaped constellations optimized for specific transmission environments. Traditional detectors often face difficulties in decoding information without pre-existing knowledge of the constellation design. This challenge can be addressed by employing DL techniques for E2E learning. Such techniques allow for the optimization of both transmitter and receiver within an AE framework, enabling them to effectively learn suitable input-to-output mappings [26]. For instance, research outlined in [38] utilizes an AE-based E2E approach to create optimal constellation designs and demapper architectures. This is particularly effective in environments with additive white Gaussian noise (AWGN) and radar interference. In [39], researchers developed an AE-based constellation design for MU interference channels, effectively addressing the issue of variable interference. The study in [9] explores bit-wise and symbol-wise transmissions in an E2E system, revealing that bit-wise AE, especially when coupled with IEEE 802.11n (Wi-Fi 4) low-density parity-check (LDPC) code, surpasses both symbol-wise AE and traditional QAM or PSK methods. The approach entails joint optimization of constellation geometry and bit labeling during the training phase of transmitter, leading to improved constellation point separation and enhanced overall performance. In [40], researchers assessed various neural demappers featuring trainable constellations in an E2E system, comparing them with conventional demapping algorithms utilizing QAM constellations in AWGN channels. The simulations highlighted that these neural demappers, with their trainable constellations, achieved a notably lower bit error rate (BER) compared to the traditional methods.

Despite the popularity and widespread adaptation of OFDM in numerous standards including the IEEE 802.11 family and fifth-generation (5G) wireless, it has certain limitations such as high peak-to-average power ratio (PAPR), cyclic prefix (CP) overhead, and pilot overhead. Such limitations and impairments from different physical layer (PHY) blocks can heavily corrupt the received signal and make the system incompatible with downstream DL tasks and applications [41]. Several studies have focused on AE-based techniques to address these limitations in OFDM. In [42], a low-complex real-valued NN was introduced to simultaneously decrease PAPR and enhance BER. [43] presented a DL-based tone reservation network aimed at reducing PAPR and expediting training, which was further advanced in [44] by integrating iterative tone reservation into a DNN layer. In addition, [45] utilized a DNN at the transmitter, combined with AE-based E2E training, to create high dimensional modulation schemes capable of managing both channel leakage ratio and PAPR.

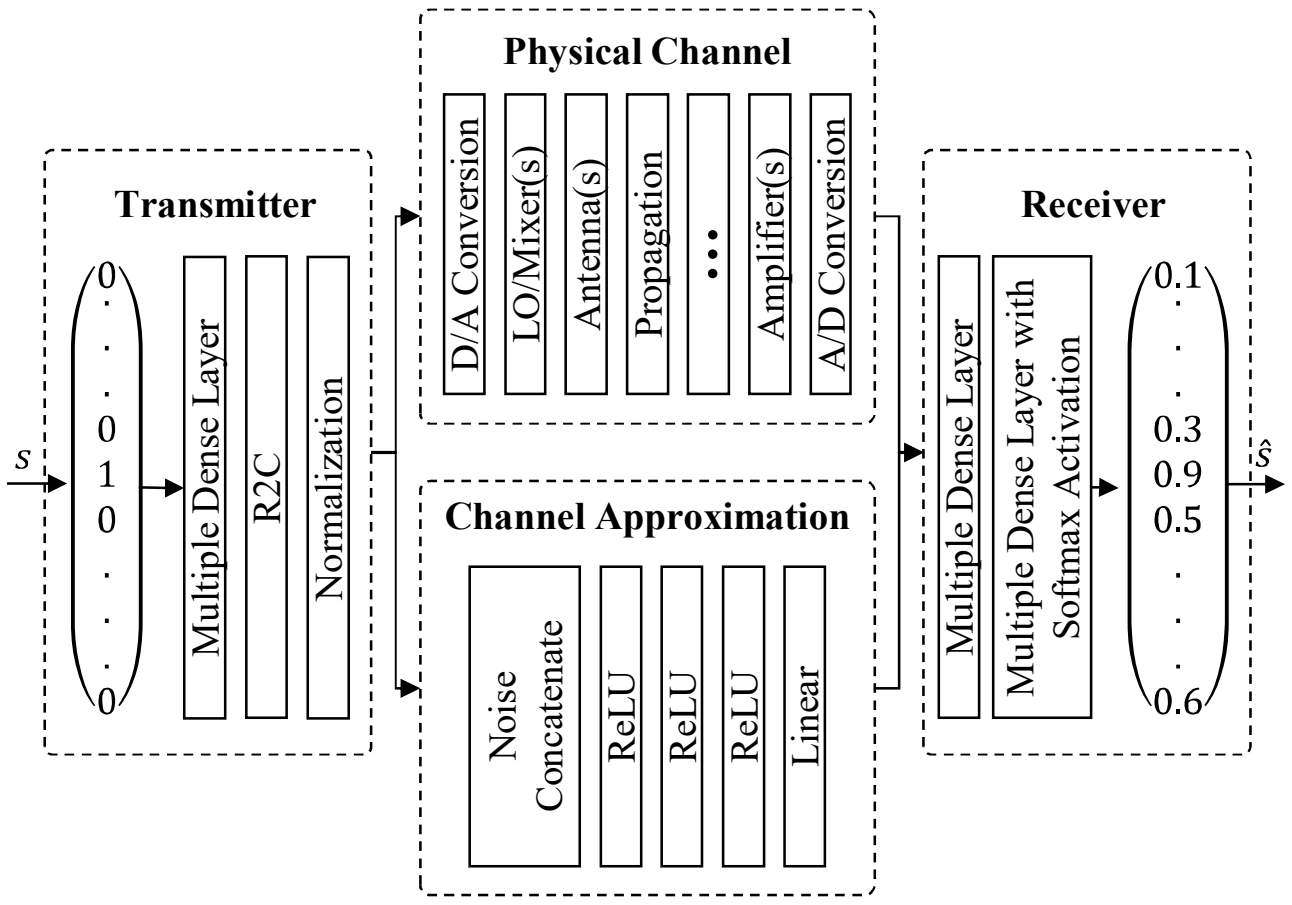

**FIGURE 6. GAN based learning system for physical channel communication. Where, $s$ is the encoded one-hot vector input, and $\hat{s}$ is the probabilistic decoded output [51].**

To enhance spectral efficiency, OFDM systems need to manage both CP and pilot overhead. In this context, [46] introduced a comprehensive E2E transceiver. This system leverages an AE-based NN combined with a trainable constellation, achieving state-of-the-art performance over practical wireless channels, all while eliminating the need for any pilot signals. The authors in [47] tackled both CP and pilot reduction challenges, demonstrating that it is possible to completely remove CP and pilots, as well as increase throughput significantly by leveraging E2E learning. [48] proposed a novel approach combining downlink pilot design and channel estimation for frequency division duplex massive MIMO-OFDM systems using NN. This approach also included a method for efficient pilot reduction, which involved progressively pruning less crucial neurons from dense layers, thereby minimizing pilot overhead and conserving time-frequency resources for data transmission. The authors in [33], [34] introduced a pilotless AE-based E2E learning method to address the channel estimation issue, implementing a CNN and two DNNs on the transmitter and receiver sides, respectively. The system in [34] proposed mini-batches of input for pilot-free communication, reducing pilot overhead.

In [49], the E2E approach in a MIMO system with multiple antennas is explored, demonstrating enhanced spatial diversity, multiplexing, and reduced BER compared to conventional MIMO systems. The work in [50] expands the single-input single-output (SISO) system from [6] to cover both SISO and MU MIMO interference systems over Rayleigh fading channels, revealing that an AE marginally outperforms a standard, non-interference, single-user MIMO system with quadrature phase shift keying (QPSK) modulation. Further in-depth analysis in [51] encompasses systems with single-antenna, multiple-antenna, and multiuser configurations. Additionally, the study integrates DL for spectrum situational awareness, covering aspects like channel modeling, estimation, signal detection, classification, and securing wireless communications. Simulation results highlight that DL based data-driven approach improves the overall performance in cases where traditional model-driven methods falter. The study showcases a GAN-based learning system for physical channel communication, depicted in Fig. 6, and explores potential of adversarial DL in executing and countering jamming attacks by sabotaging the learning process of the adversary. Furthermore, [52] introduces an AE-based E2E network utilizing recurrent neural network (RNN) to effectively handle dynamic input lengths, demonstrating its superiority over CNN-based networks.

E2E systems are characterized by their use of DNNs to learn direct mappings from transmitted messages to received signals, optimizing the entire communication process as a unified task. This fundamentally differs from traditional methods that segment the communication process into discrete encoding, transmitting, and decoding blocks, often leading to suboptimal performance due to a lack of joint optimization across these stages. E2E communication systems provide a pivotal shift in communication by shifting the focus from accurate symbol transmission to the meaning or 'semantics' of the transmitted information, thus enabling semantic communication (SemCom) applications. SemCom is designed to ensure that the intended message is understood by the receiver, regardless of the specific symbols transmitted, thereby prioritizing context and intent over traditional bit accuracy. This approach leverages the capabilities of E2E systems to adaptively learn from data, allowing SemCom to dynamically adjust to the semantic requirements of the communication scenarios [17]. For instance, in scenarios where the background knowledge of the transmitter and receiver is aligned, SemCom systems can significantly reduce the amount of data needed to be transmitted by focusing only on the information that adds new meaning or context, rather than transmitting redundant data. The relationship between E2E and SemCom systems is synergistic. While E2E systems provide a data-driven foundation by learning optimal representations and transformations of data across a communication channel, SemCom systems extend this approach by integrating semantic-level understanding. Table III compiles a detailed summary of E2E communication systems that enable data-driven PHY for wireless transmission. SemCom and its various applications enabled by E2E communication will be discussed in Section IV.

**TABLE 3. Summary of DL-based E2E communication systems for data-driven PHY**

| | Ref | Key Ideas and Findings | Major Limitations |
|---|---|---|---|
| PHY as DL-based E2E system | [6] | • Represents PHY as AE for E2E learning<br>• Adversarial network for channel capacity and CNNs for modulation classification<br>***Ch:*** *Rayleigh, AWGN;* ***Mt:*** *BER, BLER, Cross-Entropy Loss, Classification Probabilities;* ***DL:*** *AE, radio transformer networks (RTN), Adversarial, CNN* | • Not scalable to long block lengths<br>• Consider communications as a classification problem – not suitable for alternate data output<br>• Assumes transfer function of the channel – not practical for real channels/hardware |
| | [22] | • Built E2E communications system prototype with two SDRs<br>• Entire physical layer processing carried out by NN<br>***Ch:*** *AWGN, Over the air;* ***Mt:*** *BER;* ***DL:*** *AE, RL* | • BLER performance is about 1 dB worse than baseline system |
| | [25] | • Stochastic perturbation to train E2E without relying on explicit channel models<br>• Achieves comparable BER with [6] even with practical channel<br>***Ch:*** *AWGN, Rayleigh Block Fading, Over the air;* ***Mt:*** *BER;* ***DL:*** *AE* | • Requires more training epochs to converge compared to [6] |
| | [26] | • Implements learning of communications systems over any channel without prior assumptions<br>• Alternate training between supervised and RL-based at the receiver and transmitter<br>***Ch:*** *AWGN, Rayleigh, Block Fading;* ***Mt:*** *BLER, PSNR, SER, Cross-Entropy Loss;* ***DL:*** *AE, RL* | • Additional reliable channel required for feedback<br>• Transmitter training does not scale with channel when using proposed gradient approximation<br>• Long coherence time at training |
| | [28] | • Proposed alternating training of [26] with noisy feedback<br>• Designed system that can learns to transmit over unknown channel without feedback link<br>***Ch:*** *AWGN, Rayleigh;* ***Mt:*** *BLER, MSE;* ***DL:*** *AE, RL* | • Higher BLER compared to Agrell modulation [53] scheme over AWGN channel |
| | [30] | • Proposed conditional GAN to represent channel environment that is differentiable<br>• Proposed CNN to address dimensionality problem<br>***Ch:*** *AWGN, Rayleigh Fading, Frequency-Selective Multipath, WINNER II;* ***Mt:*** *BER, BLER;* ***DL:*** *AE, GAN, CNN* | • Higher BLER than QAM over Rayleigh fading<br>• Higher BER and BLER in SNR over WINNER II<br>• Training instability, vanishing gradient, and slow convergence |
| | [31] | • Proposed differentiable RA-GAN for channel environment<br>• Constructed loss function of RA-GAN to solve the overfilling problem in conventional GAN<br>• Generated realistic received signal, and achieved superior performance compared to [29]<br>***Ch:*** *AWGN, Rayleigh, Deep MIMO Ray tracing, Optical Fiber;* ***Mt:*** *BLER;* ***DL:*** *AE ,GAN* | • Computational complexity for residual generator<br>• Performance gain is negligible compared to RL [26] and WGAN [54] based training over AWGN |
| | [33] | • Pilot-free E2E system with wireless channel modelled as stochastic convolutional layer<br>• Designed feature extraction module for the channel using bilinear production<br>***Ch:*** *Frequency-Selective, MIMO Channels;* ***Mt:*** *BER, MS-SSIM;* ***DL:*** *AE ,DNN, CNN* | • Not adaptive to different channel and SNR values |
| | [35] | • Implemented E2E with DNN-based channel model, where transmitter and receiver are composed with one dimensional CNN and can input infinite-length sequence<br>• Compared to [30], faster convergence, better stability with lesser number of parameters, no model collapse issue, and requires only single network<br>***Ch:*** *Frequency-Selective Multipath, Rayleigh Channels;* ***Mt:*** *BER, BLER, Loss, KL Divergence;* ***DL:*** *AE ,DNN, CNN* | • Baseline outperforms BLER performance over Rayleigh channel<br>• Does not use real dataset |
| DL-based E2E in Communication Standards | [36] | • Extended E2E system proposed in [22] to OFDM with CP<br>• Achieved single-tap equalization (no need for explicit pilots or equalization)<br>• Robust against impairments and errors from hardware and sampling synchronization<br>***Ch:*** *AWGN, Multipath-Fading, Frequency Selective Fading;* ***Mt:*** *BLER;* ***DL:*** *AE, RTN* | • Does not achieve diversity gain in fading channels as it uses RTN-based AE per single sub-carrier<br>• Use of RTN for CFO compensation is not shown<br>• Short length sub-carrier messages |
| | [37] | • Novel single-user MC-AE based system with modulation and demodulation blocks as DNNs<br>• Unlike [6] and [22], CSI and received signal is directly fed to decoder without any domain knowledge of channel equalizer<br>• Extended to multi-user scenario, with two novel DNN structures for uplink and downlink, and novel loss function for fairness and fast convergence<br>***Ch:*** *AWGN, Rayleigh Fading Channel;* ***Mt:*** *BLER loss;* ***DL:*** *AE, DNN* | • Performance of multi-user MC degrades with overloaded transmission, especially in uplink<br>• Improvement in convergence rate and fairness by the proposed novel loss function is not well understood and explained<br>• High computational for decoding complexity |
| | [38] | • Designed finite size constellations and corresponding soft demapping rules which is jointly optimization using AE<br>• Proposed designs are efficient with both coded and uncoded data<br>• Introduces new demapping rule that is adapted to the additive radar interference channel<br>***Ch:*** *AWGN Channel with Additive Radar Interference;* ***Mt:*** *SER, BER, Rate in Bits;* ***DL:*** *AE* | • Adaption of the system to different transmission medium is not considered<br>• Rely on only a fixed training model with offline optimization |
| | [39] | • Proposed adaptive deep learning (ADL) for robust link with m-user interference channel<br>• Studied learned constellation and how it is affected by multi-user interference<br>***Ch:*** *AWGN;* ***Mt:*** *BER, SER;* ***DL:*** *ADL, AE* | • For single use, no significant improvement between AE and conventional BPSK/QPSK<br>• Complex and real channel models not considered |
| | [9] | • Fully differentiable neural iterative demapping and decoding (IDD) with significant gains<br>• Implements on SDR, and training of the E2E over actual channel<br>• Training on the BMI allows integration with practical BMD receivers<br>***Ch:*** *AWGN, Over the Air;* ***Mt:*** *BER;* ***DL:*** *AE, RL* | • Baseline outperforms symbol-wise AE<br>• Heuristic labelling (symbol-wise AE) not optimal |
| | [40] | • Performance evaluation of neural demapper on trainable constellation using AE and AE-RL<br>• Considered constellation design corresponding to 16-QAM to 1024-QAM<br>***Ch:*** *AWGN;* ***Mt:*** *BER;* ***DL:*** *AE, RL* | • Improvement with AE is negligible at lower bit per symbol constellation<br>• Does not consider complex and fading channels |
| | [45] | • NN transmitter and receiver for high dimensional modulation, and demapping task<br>• E2E optimization to control PAPR and ACLR (both differentiable)<br>***Ch:*** *AWGN;* ***Mt:*** *Rate, ccdf, PSD;* ***DL:*** *AE* | • Did not consider realistic and complex channel<br>• Few subcarriers implemented in the system |
| | [46] | • Implemented E2E over time and frequency selective fading channel using OFDM<br>• OFDM with large subcarriers and, eliminates orthogonal or superimposed pilots entirely<br>• High throughput and reliability with pilotless transmission, and consider PAPR constraints<br>***Ch:*** *Time-Selective, Frequency-Selective;* ***Mt:*** *BER, Goodput;* ***DL:*** *AE* | • Performance of QAMs with iterative and non-iterative is not shown for comparison<br>• Though pilotless communication system is evaluated, cyclic prefix is still used in the system |
| | [48] | • NN based joint pilot design and downlink channel estimation in FDD Massive MIMO OFDM<br>• Pilot reduction using pruning technique of NN<br>***Ch:*** *Multipath MIMO, COST 2100 [55];* ***Mt:*** *NMSE, SER;* ***DL:*** *AE, CNN* | • Needs to be trained at different SNR values<br>• High computational complexity |
| | [51] | • Propose E2E system for multiuser, single-antenna and multiple antenna communication<br>• Implements DL in spectrum situation awareness such as channel estimation, channel modelling, signal classification and signal detection<br>• Implements wireless security system utilizing GAN<br>***Ch:*** *Multiuser MIMO Interference Channel;* ***Mt:*** *BLER, SER, Throughput, Loss, Defense Budget;* ***DL:*** *AE, DNN, GAN, CNN, RTN* | • High fluctuation between the discriminator and generator loss in GAN<br>• Delayed convergence by GAN, and does not converge fully<br>• To obtain best defense high iterations are required |
| | [52] | • Proposed RNN based communication system with E2E transceiver optimization<br>• Does not have any limit to the length of input sequence with the help of CNN and LSTM<br>***Ch:*** *AWGN, Multipath Fading Channels;* ***Mt:*** *BER;* ***DL:*** *AE, LSTM, RNN, CNN* | • Not robust over fading channels as it does not use block AEs<br>• Uses low Code rate- high redundant information<br>• Has low spectrum efficiency |

Notation: ***Ch*** are the channel models used, ***Mt*** are evaluation metrics, and ***DL*** are DL algorithms used in the experiment.

## IV. SEMANTIC COMMUNICATION SYSTEM

Semantic communication (SemCom) promises a paradigm shift in data transmission, moving from traditional bit-focused approaches to understanding the essence of the transmitted information— the semantics. This system has the capacity to interpret and convey the semantic content of a message, leveraging the shared background knowledge between transmitter and receiver to minimize semantic errors and reduce the need for transmission symbols. At the core of this transformative approach lies end-to-end (E2E) learning and optimization, which transforms communication systems into deep neural network (DNN) architectures. It is capable of learning optimal representations for tailoring the encoding and decoding processes to the specific characteristics of the channel and the nature of the data being conveyed. Various semantic applications have been realized to encapsulate and convey the semantics effectively across different modalities such as text, image, video, audio and multimodal transmission over multicarrier waveforms. For instance, in semantic text transmission E2E learning utilizes recurrent neural networks (RNNs) and transformer models to capture and convey complex linguistic structures. For images and video, convolutional neural networks (CNNs) have been instrumental in advancing the efficacy of transmission, outperforming standard compression methods. Similarly, semantic multimodal communication systems have utilized more complex DNN architectures. These applications are driven by E2E learning frameworks that integrate and interpret semantic information based on shared knowledge, optimizing the message for clarity and conciseness across diverse physical channels. Utilizing E2E optimization, SemCom not only revolutionizes data transmission by compressing information to its semantic core but also ensures the fidelity of the intended meaning during transmission, regardless of channel conditions or potential distortions. This marks a significant advancement in the evolution of communication technologies. In this section, we discuss SemCom and review various semantic applications that utilize E2E optimization to enable SemCom across wireless network environments.

### A. SEMANTIC COMMUNICATION PARADIGMS

In this section, we delve into the universal paradigms that broaden the scope of SemCom. These paradigms leverage sophisticated data structures and communication theories, incorporating broader contexts and objectives that surpass traditional data transmission methods.

**Probabilistic graphs**: Probabilistic graphs, including Bayesian networks and Markov models, provide a robust framework for modeling uncertainties and dependencies in communication systems. In SemCom, these models are crucial for predicting and interpreting the semantics of the information transmitted, which is particularly useful in dynamic environments where prior probabilities and the relationships between data points are subject to change. For instance, in a wireless network system, probabilistic graphs can be employed to model and predict the flow of semantically relevant information, thereby optimizing the communication channels for context-aware data transmission.

**Knowledge graphs**: Knowledge graphs represent complex relationships between entities and are instrumental in enhancing semantic comprehension. In SemCom, knowledge graphs are utilized to encode not just data, but also the interrelations and attributes that define their semantic context. This paradigm enables communication systems to process and interpret data within a rich semantic framework, significantly improving the efficacy of information retrieval and decision-making processes. By leveraging ontologies and metadata, knowledge graphs facilitate a deeper understanding of the data, enhancing both the precision and relevance of communication in systems such as intelligent personal assistants and advanced recommender systems.

**Semantic-oriented Communication**: Focusing on the relevance of the content itself, semantic-oriented communication emphasizes transmitting data that is semantically significant to the receiver rather than merely optimizing the data transmission rates. This approach aligns the transmitted data with the semantic expectations and needs of the recipient, ensuring that the communication is not only efficient but also contextually relevant. For example, in a medical data exchange scenario, semantic-oriented communication would prioritize the accuracy and relevance of patient data transmitted between healthcare providers, ensuring that the semantics of the medical conditions are preserved and clearly understood.

**Goal-oriented communication**: Goal-oriented communication aligns the transmission process with specific outcomes or objectives, utilizing semantics to ensure that the information transmitted directly contributes to achieving predefined goals. This paradigm is particularly useful in collaborative environments like collaborative robotics and Internet of Things (IoT) applications where the communication must trigger specific actions or responses based on the semantic understanding of the data received. It addresses the effectiveness problem of communication systems by ensuring that every bit of the transmitted data serves a functional purpose and directly aids in accomplishing the task at hand.

**Semantic-aware communication**: This paradigm extends the concept of SemCom by integrating semantic understanding directly into the communication protocols, enabling systems to be aware of the semantic significance of the information being processed and transmitted. Semantic-aware communication systems are capable of dynamically adjusting their behavior based on the semantic interpretation of incoming data, which is crucial for applications involving complex decision-making processes, such as autonomous driving and real-time data analysis systems.

While the field of SemCom encompasses a wide range of paradigms and approaches, this article specifically focuses on how deep learning (DL) can transform the physical layer (PHY) of communication systems and enabling efficient and effective semantic interpretation and transmission across various modalities (text, image, audio, video, and multimodal). Consequently, this paper categorizes SemCom by various application modalities. The other universal SemCom paradigms for wireless communication systems are out of scope and can be referred to in [12], [13], [14].

### B. QUANTIFYING SEMANTIC INFORMATION

Semantic entropy is the measure of *semantics,* and it is fundamental to developing systems with semantic information. Unlike Shannon's entropy which considers the probability distribution of only random variables, semantic entropy considers the structures of elements in the set of random variables with logical relations [12]. Semantic entropy was first conceptualized in the language system and the measure of semantic entropy of a sentence is given as:

$$H(s,e) = -log c(s,e), \tag{9}$$

where, $s$ is the original sentence and $c(s,e)$ is the degree of confirmation of the sentence $s$ on evidence $e$, given as:

$$(s,e) = \frac{m(e,s)}{m(e)}. \tag{10}$$

$m(e)$ and $m(e,s)$ are the measure of natural language inference (logical probability) of the sentence $s$ on evidence $e$, and on that of evidence $e$ and itself, respectively. Besides probabilistic model, semantic entropy can also be measured using comprehension model that utilizes the latent background knowledge [12]. Entropy measure at time step $t$ is given by:

$$H(t) = -\sum_{v_M \in \mathcal{V}_{\mathcal{M}}} P(v_M|v_t) log P(v_M|v_t), \tag{11}$$

where, $\mathcal{M}$ is the set of models, $P(v_M|v_t)$ is the conditional probability of $v_M$ given $v_t$ , and $\mathcal{V}_{\mathcal{M}} = \{v_M | v_M(i) = 1\ iff\ M_i = M, and\ M\ is\ a\ unique\ model\ in\ \mathcal{M}\}$ . Similarly for translation task, the semantic entropy of each word is given by:

$$H(w) = H(T|w) + N(w),$$
$$= -\sum_{t \in T} P(t|w) log P(t|w) + P(NULL|w) log F(w) \quad (12)$$

where, $w$ is the original word, $T$ is the set target words, $N(w)$ is the untranslatable word, $F(w)$ is the frequency of $w$, and $H(T|w)$ is the translational inconsistency. Furthermore, semantic entropy for classification task can be measured using membership degree in axiomatic fuzzy set theory [12]. The matching degree $D_j(\zeta)$ characterizing semantic entropy for data in class $D_j, j \in \{1,2,\ldots,m\}$, is given by:

$$D_j(\zeta) = \frac{\sum_{x \in X_{C_j}} \mu_\zeta(x)}{\sum_{x \in X} \mu_\zeta(x)}, \quad (13)$$

where, $\zeta$ is the semantic concept and $\mu_\zeta(x)$ is the membership degree of data sample $x$. For the elements in class $C_j$, the semantic entropy is given by:

$$SE_{C_j}(\zeta) = -D_j(\zeta) log_2 D_j(\zeta). \quad (14)$$

Next, the semantic entropy of concepts $\zeta$ on X is defined by:

$$SE_\zeta = \sum_{j=1}^{m} SE_{C_j}(\zeta). \quad (15)$$

The above semantic entropy definitions are for single tasks only. Semantic information for any given task and source can be quantified by defining the minimum number of semantic queries about data $S$, and the answer to the queries used to predict task $V$ [12], which is given by:

$$H_Q(S;V) \triangleq \min_E E_s\left[\left|Code_Q^E(S)\right|\right]$$
$$such\ that\ P\left(v \middle| Code_Q^E(S)\right) = P(v|s), \forall s, v. \quad (16)$$

$Code_Q^E(S)$ is the query vector extracted from $S$ with the semantic encoder $E$. Although this method is suitable for semantic entropy on any given task and source, finding an optimum semantic encoder is as challenging as finding semantic entropy. To summarize, unlike unified Shannon entropy, the current definitions of semantic entropy are diverse, and specific to a given task only. Therefore, a unified semantic entropy definition is still an open research challenge. The aforementioned semantic entropy definitions are referred to in [12], which also discusses the information theoretic fundamentals of semantic and task-oriented communication in detail.

Furthermore, there have been recent theoretical advances in semantic information that offer significant insights into the quantification and understanding of semantic aspects in communication systems. The concept of synonymous mapping [56] for the construction of a semantic framework and the introduction of semantic entropy in channel coding [57] provide a theoretical foundation for new communication paradigms that transcend the limitations of Shannon's entropy by considering the semantic significance of the transmitted messages. Authors in [56] propose a structured framework for semantic information theory (SIT), which extends classic information theory by introducing semantic-related concepts and measures. It introduces the concept of synonymous mapping, where semantic information is mapped onto structural information, capturing the synonymous nature of semantic expressions. It also develops measures such as semantic entropy $Hs(U\sim)$, semantic mutual information $Is(X\sim;Y\sim)$, and semantic capacity $Cs$, which are specific to semantic contexts. Additionally, the paper extends classical coding theorems to the semantic domain, asserting that semantic capacity $Cs$ may exceed classical channel capacity and semantic rate-distortion function $Rs(D)$ can be lower than classical limits, potentially leading to more efficient coding schemes. Authors in [57] address semantic channel coding, focusing on scenarios where multiple source sequences can represent the same semantic meaning (many-to-one source mappings). This paper proposes a new definition of semantic entropy as the uncertainty in the semantic interpretation of symbols within a knowledge base (KB) context, effectively linking it with existing semantic entropy definitions under certain conditions. It establishes a theorem for SemCom with many-to-one sources, demonstrating that SemCom can transmit semantic information accurately even under non-zero bit error rates. The paper also proves both the feasibility of the proposed semantic coding theorem using joint probability tools and the converse using a generalized Fano's inequality. These works collectively advance the understanding of semantic information theory, suggesting that SemCom could surpass traditional Shannon-based limits by incorporating semantic context and synonymy in the encoding and decoding processes. To clarify the evolution and application of semantic entropy, we present a comparative analysis of the newly proposed frameworks in the following sections.

### C. TEXT APPLICATIONS

In separate source-channel coding in PHY, data compression (source coding) is followed by the addition of redundancy (channel coding) to ensure the integrity of transmitted data over the channel. This approach is optimal for transmission over discrete memoryless channels with infinite block lengths but falters when encoding documents of finite length. The authors in [58] were the first to propose DeepJSCC, a DNN-based joint source-channel coding (JSCC) scheme for structured data transmission such as natural language over a noisy channel, as shown in Fig. 7. Utilizing RNNs for both encoding and decoding, the system maps sentence into a semantic space where semantically similar sentences are positioned closely together. This structure emphasizes the preservation of semantic content over strict bit-level accuracy. Comparative results showed that the DL system, especially under tight bit budget constraints, achieved a lower word error rate (WER) when using an erasure channel with Reed-Solomon (RS) codes for channel coding, compared to three benchmark source coding techniques (gzip, Huffman, and fixed length coding).

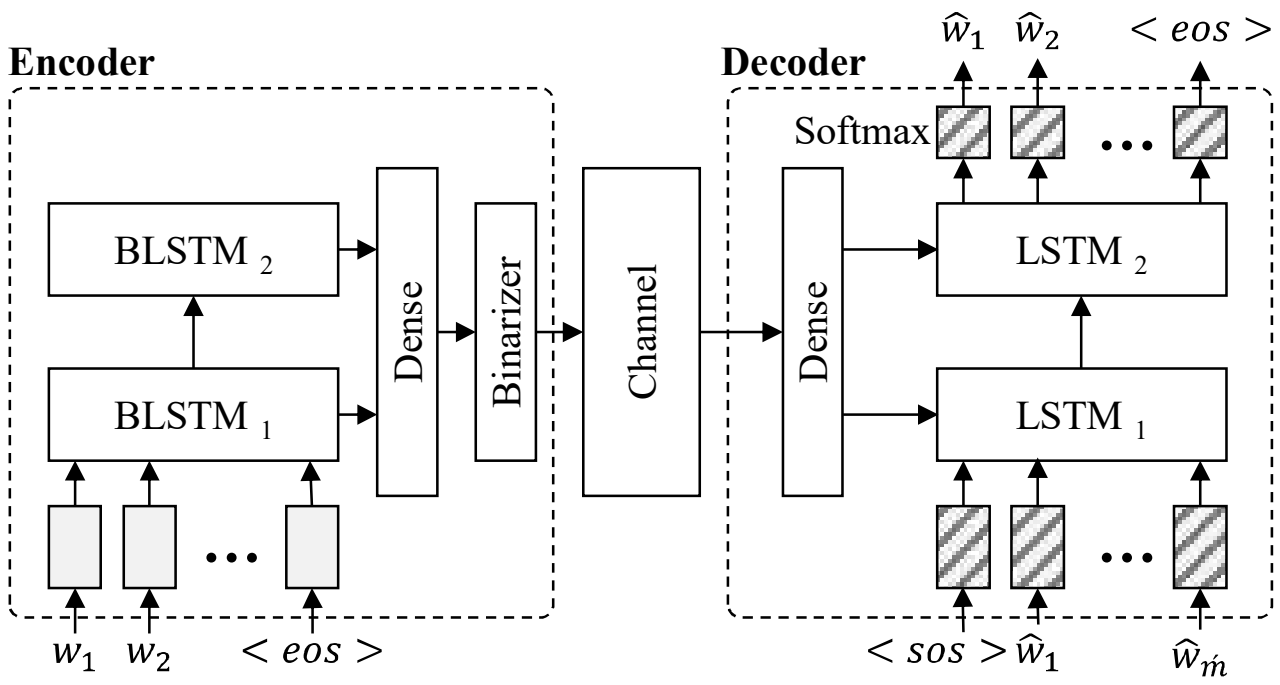

**FIGURE 7. E2E optimization using encoder-decoder architecture for text transmission. $[w_1, \ldots, w_m]$ is the set of input words, $[\hat{w}_1, \ldots, \hat{w}_{\hat{m}}]$ is the set of output words, and $< sos >$ and $< eos >$ are start and end of a special sentence word, respectively [58].**

The authors in [59] proposed an E2E communication system for text transmission, known as deep semantic communication (DeepSC). Contrary to traditional communication systems that prioritize bit or symbol error correction, DeepSC focuses on preserving the semantic integrity of sentences, ensuring that the transmitted content retains its original meaning. The framework, grounded in the transformer architecture, integrates both semantic and channel encoding/decoding processes to counter both channel noise and semantic distortion, respectively. Fig. 8(a) shows the proposed framework of DeepSC and Fig. 8(b) illustrates the E2E DL model architecture. To ensure an accurate understanding of semantics while optimizing system capacity, the receiver employs a dual loss function strategy comprising cross-entropy and mutual information. Furthermore, to better understand the semantic performance of the system, a novel metric known as 'sentence similarity' is introduced. This metric is designed to mirror human judgment more closely than traditional metrics, offering a more nuanced understanding of SemCom performance. To adapt the system to various communication scenarios, transfer learning is introduced, allowing the DeepSC model to leverage weights from pre-existing models. Experiment results showed DeepSC outperforms all traditional communication systems under additive white Gaussian noise (AWGN), Rayleigh fading and Rician fading channels, particularly in low signal-to-noise ratio (SNR) regions. Furthermore, it achieves rapid convergence and adaptability to diverse channels due to transfer learning. However, the performance of DeepSC may degrade in highly dynamic environments where channel conditions change rapidly, as the system requires stable channel conditions during short timeframes for optimal performance. Further improvement could be made by enhancing the ability of the model to adjust more quickly to these rapid changes.

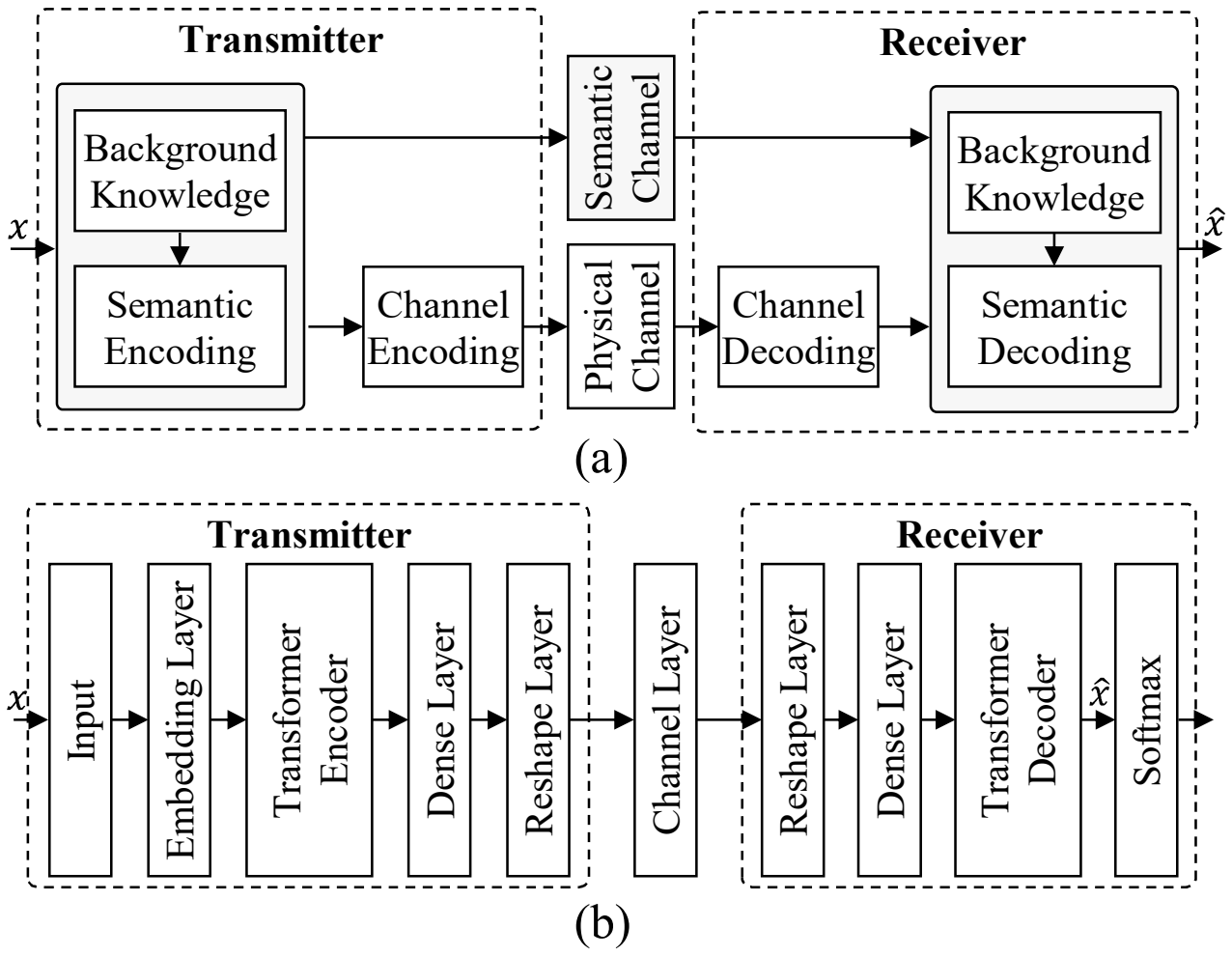

**FIGURE 8. E2E optimization for text transmission. (a) is DL based SemCom framework for text transmission, and (b) is E2E optimization using encoder-decoder architecture for text transmission. Where, $x$ and $\hat{x}$ are the input and output vectors, respectively [59].**

The authors in [60] extended DeepSC to L-DeepSC, a DL-based lightweight distributed SemCom designed for text-data transmission. Considering the unsuitability of resource-constrained IoT devices for DL computations, the study explores a network that integrates edge/cloud platforms with IoT devices. The platforms handle DL computations, while the IoT devices are responsible for gathering data and managing transmissions essential for model training. The authors examined the effects of fading channels on L-DeepSC training and introduced a channel state information (CSI) assisted method to counter distortions during data transmission using a deep de-noise network. They also introduced a semantic constellation optimized for low-capacity IoT devices, ensuring cost-effective data communication. Additionally, L-DeepSC circumvents over-parametrization by pruning model redundancy and reducing weight resolution. This ensures semantic level communication efficiency, making the system viable for IoT devices and concurrently minimizing the bandwidth needed for transmitting model weights between these devices and their cloud/edge counterparts. The system demonstrates adaptability under power and latency constraints, with CSI proving effective during training over fading channels. It also showcases the benefits of a finite-bit constellation for cost-efficient IoT device implementation. Notably, in low SNR regions, the system maintains competitive performance alongside a significant 40-fold compression ratio. One limitation of the L-DeepSC is the potential for decreased performance in environments with extremely limited network connectivity, where even minimal data transmissions can be challenging. Enhancements in network resilience and data encoding could further optimize performance in such constrained scenarios.

Drawing inspiration from [59], [60] the authors in [61] unveiled a novel SemCom tailored for text transmission based on a universal transformer (UT). This approach effectively tackles a notable shortfall in traditional transformer-based SemCom systems—their inability to adjust to the diverse semantic nuances within different sentences and to accommodate fluctuating channel conditions. Previous systems often deployed a fixed transformer structure for natural language processing (NLP) tasks, thus ignoring the nuances and differences in semantic information from one sentence to another. In contrast, the authors in [61] study UT paired with an adaptive circulation mechanism, which allows the system to dynamically adjust its processing cycles based on the unique semantic characteristics of each sentence, making it inherently flexible. This adaptability ensures the system can deliver optimized E2E performance across different channel conditions, notably enhancing its robustness, particularly in environments with fluctuating SNR. Experimental results show that UT outperforms conventional transformer based SemCom. However, the computational complexity of UT can still pose challenges in real-time applications, where processing speed is critical. Future research could explore methods to streamline the processing steps within UT to enhance efficiency without compromising the adaptability of the system.

The authors in [62] introduce SemanticRL, a reinforcement learning (RL)-based SemCom that focuses on *semantic blindness* in common objective functions, as opposed to traditional methods that focus on network design. Acknowledging the challenge of non-differentiable semantic metrics and the instability due to noisy channels, the study introduces a self-critical RL approach. This method facilitates effective learning across various user-defined semantic metrics, effectively navigating the complexities of SemCom. The method is integrated with a self-critic stochastic iterative updating (SCSIU) training mechanism, specifically for crafting an independent semantic transceiver. The proposed method was tested on a European-parliament dataset, showcasing its superior performance in extracting semantic meaning and handling semantic noise. The paper further highlights the robust generalizability of the method through a RL-based image transmission extension. It introduces the SemanticRLJSCC solution, which emphasizes semantic similarity over bit-level accuracy, demonstrating significant improvements in semantic metrics. By integrating self-critic training into a decoupled transceiver, the study provides a full-scale solution for wireless SemCom, addressing challenges of non-differentiability in both channels and objectives. However, the reliance on reinforcement learning may introduce complexities in training and tuning, particularly in dynamic environments where channel conditions are variable. Further development could explore simplifying the training process or enhancing stability in fluctuating conditions.

The authors in [63] introduce a novel semantics-aware communication framework that emphasizes the importance of semantic encoding and the broader concept of SemCom. While current semantics-aware communication techniques offer some solutions, they exhibit notable limitations, such as oversimplifying the dynamic channel state by modeling it as a static AWGN— a communication problem. Additionally, there is an over-reliance on NLP models for semantic representation, which does not fully address the broader objective of semantic transmission, rendering these techniques unsuitable for widespread applications— a semantic problem. Addressing these challenges, the article introduces a joint semantics-noise coding (JSNC) mechanism, which adapts the depth of semantic representations based on both the channel state and sentence semantics. Moreover, a novel SemCom is introduced, one that is rooted in learning from semantic similarities rather than traditional metrics such as cross-entropy loss or mean squared error (MSE), allowing for greater flexibility, including working with non-differentiable semantic metrics. Experimental results show that at a certain point, the proposed method achieved 24% lower WER compared to the transformer baseline which is significantly lower than conventional methods such as RS coding. However, the reliance of the system on adaptive representations could lead to computational inefficiencies, particularly when deployed in environments with limited processing capabilities. Future work could aim to optimize these adaptive mechanisms to reduce computational demands while maintaining system performance.

In [64], the authors present SC-RS-HARQ, a fusion of semantic coding (SC) with RS channel coding, augmented by the hybrid automatic repeat request (HARQ) to significantly diminish errors in SemCom. SC adapts the code length to the sentence length, outperforming the traditional fixed-length code by reducing word transmission errors, especially in environments with high bit error rate (BER). The study further introduces an E2E design to enable SemCom called SCHARQ, which is a transformer-based system that enhances transmission efficiency by transmitting incremental bits in accordance with HARQ. SCHARQ dynamically adjusts its code length based on the receiver requirements, offering enhanced performance compared to traditional RNN and transformer-based methods. This adaptability proves particularly effective in handling diverse sentence lengths and in high BER conditions, ensuring efficient and reliable communication. Additionally, the study introduces Sim32, an error detection network focused on identifying meaning-errors in received sentences. This allows sentences with minor errors to be accepted, given their overall semantic meaning remains consistent, thus conserving transmission resources. Experimental results show that the proposed parallel SC-RS-HARQ achieves superior performance compared to traditional systems. Furthermore, Sim32 with a cyclic redundancy check improves the semantic coder for semantic error detection. While these innovations mark significant advances, the complexity of integrating semantic detection with traditional error-checking could pose implementation challenges. Enhancements in the integration process or further refinement of the semantic error detection capabilities could foster broader adoption and enhance the robustness of the system.

The authors in [65] introduce a SemCom framework for textual data transmission over a network. Within this framework, a base station (BS) extracts the semantic features of the textual data, representing it as knowledge graph (KG) formed by a set of semantic triples, before transmitting it to individual users. These users regenerate the initial text from the semantic information using a graph-to-text generative model. The authors suggested a new metric named the metric of semantic similarity (MSS), aiming to evaluate the semantic accuracy and integrity by comparing the reconstructed text with the original. Considering the limitations of resource-constrained devices and the need to manage transmission delays, the BS is tasked with selectively allocating resource blocks to each user while also determining the optimal portion of semantic information for transmission. This strategy introduces an optimization challenge, focused on enhancing the overall MSS by simultaneously fine-tuning resource allocation and the selection of semantic information for transmission. To address this complex issue, the authors introduce the attention-based proximal policy optimization (APPO) algorithm, specifically designed to manage resource distribution and the transmission of semantic information efficiently. The algorithm leverages an attention network to recognize the significance of each semantic triple, building correlations between the distribution of these triples in the semantic data and the overall MSS. Furthermore, the APPO algorithm showcases an inherent capability to self-modify its learning rate, ensuring it converges to a locally optimal resolution. The simulation results show the effectiveness of this approach, achieving a substantial reduction of 41.3% in data transmission from the BS, while doubling the total MSS, outperforming traditional non-SemCom systems. Nevertheless, the complexity and computational demands of managing dynamic resource allocation while maintaining high semantic integrity pose significant challenges, particularly in real-time applications. Future developments could focus on streamlining these processes or developing more efficient algorithms to handle dynamic changes more effectively.

The study in [66] presents an innovative encoding method based on part-of-speech semantics, designed to shorten codeword lengths and increase the semantic distinction between words sharing the same codeword. This approach simplifies the process of message recovery, making it more direct and intuitive. At the decoder, the study employs the N-gram language model alongside a unique context-based dynamic programming algorithm. Two distinct learning models, the continuous bag-of-words (CBOW) and a long short-term memory (LSTM) based model, are utilized to extract contextual correlations forming the background knowledge vital for decoding. Using seven metrics, the authors evaluate their methods based on codeword length (effectiveness) and semantic similarity (reliability). Experimental results indicate that the system not only improves semantic accuracy across varying communication channels but also reduces the bit requirements for message transmission. Among the two semantic feature extraction techniques, the LSTM-based model achieved better performance compared to CBOW. Furthermore, the approach ensures that even as the number of words increases, the complexity of the encoding and decoding processes remains stable, as long as the dimensions of the context and feature windows remain constant. This constancy preserves the system efficiency regardless of the text length. However, the practical implementation of this system may face scalability issues as the complexity and computational demands can grow significantly with larger vocabulary sizes and more complex sentence structures. Future research could explore optimization techniques to manage these challenges more effectively, particularly for real-time applications.

In [67], the authors present a novel UDSem, a unified distributed learning framework for E2E SemCom over wireless networks. Traditional SemCom requires a vast amount of data, which leads to privacy concerns, regulatory concerns, and potential network congestion. To circumvent these challenges, the authors use a federated learning (FL)-based algorithm to design UDSem. This framework allows for devices to locally compute gradients based on their data and then contribute to a global update, ensuring both efficient training and data privacy. UDSem stands out for its flexible learning mechanism, tailored to the computational capabilities of each device. It segments the system into numerous modules that independently update parameters on both client devices and the central server. Additionally, UDSem employs a global aggregation method, ensuring neural parameters are updated cohesively across all modules, thus optimizing both individual device performance and overall system efficacy. Unlike traditional approaches like split learning, UDSem splits the neural modules of SemCom, trains them separately on each device, that consequently reduces communication overhead of sending numerous neural parameters to a central server. Experimental results on two public datasets for text and images showed an enhanced convergence rate and semantic interpretation compared to other methods. Nevertheless, UDSem still encounters challenges related to algorithmic stability and the consistency of updates across heterogeneous devices. Addressing these could further refine the system efficiency and robustness in diverse network conditions.

In [68], the authors delve into and clarify the primary performance constraints of [59] in scenarios involving both single and multiple sources of interference, offering a detailed examination of how such conditions impact system efficiency. While SemCom aims to reduce power usage, bandwidth consumption, and transmission delay by transmitting only semantically relevant information, the efficiency of SemCom can be compromised by radio frequency interference (RFI), leading to significant semantic noise. The authors present a probabilistic framework suggesting that DeepSC [59] might produce semantically irrelevant sentences under significant RFI, a hypothesis that was corroborated through Monte Carlo simulations. The study emphasizes the limitations of SemCom techniques like DeepSC [59], especially their susceptibility to RFI-induced wireless attacks. It critically assesses the widely held belief that SemCom performs well in low SNR conditions, challenging this assumption with insightful observations. Additionally, the study has laid the groundwork for the future design of interference-resistant and robust (IR2) SemCom for 6G by emphasizing the need for real-time robust RFI detection and mitigation, both at the technical and semantic levels. To advance this area, further development could focus on integrating adaptive interference mitigation strategies that dynamically adjust to the variability and unpredictability of real-world wireless environments.

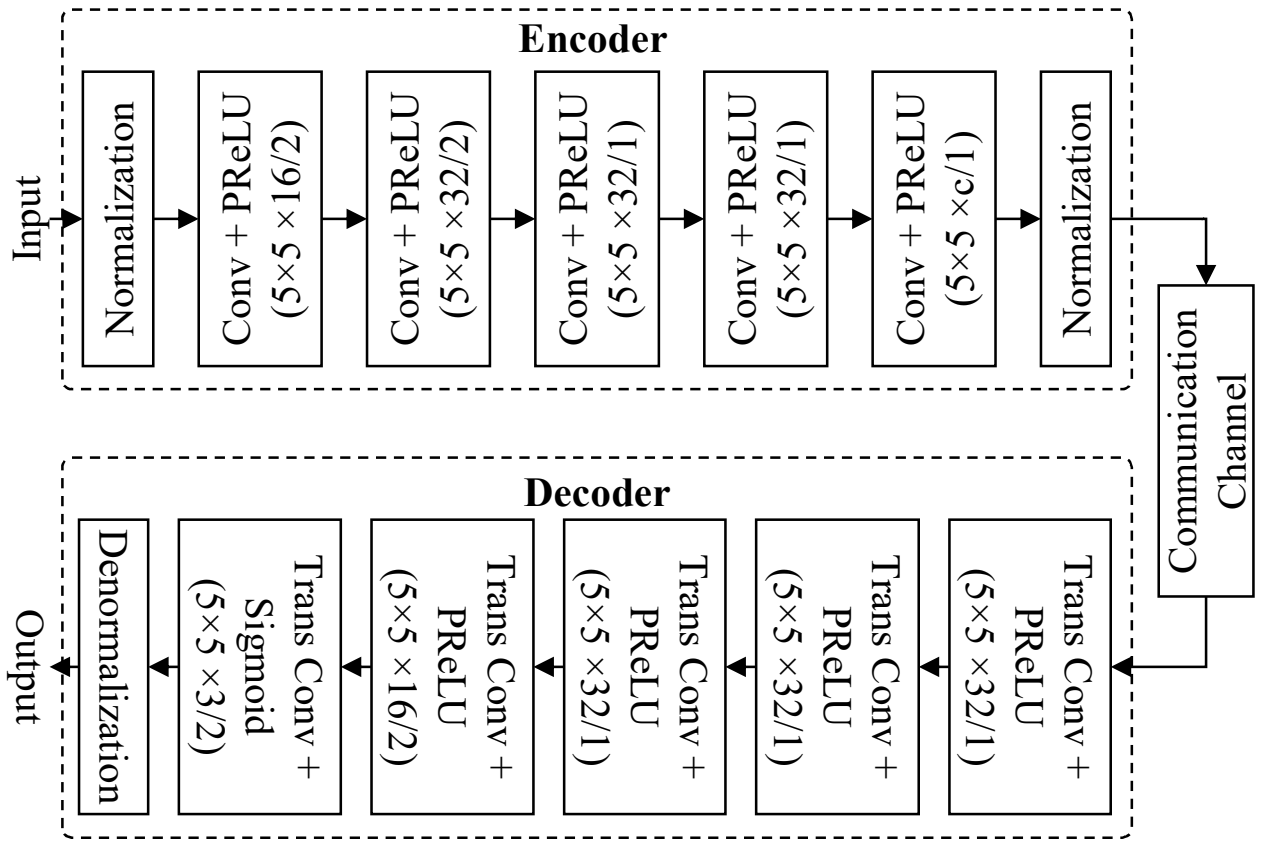

**FIGURE 9. AE based DeepJSCC structure for image transmission with E2E optimization [69].**

### D. IMAGE APPLICATIONS

Building on the principles of text-based DeepJSCC in [58], [69] explores its application in semantic image transmission, enhancing the model with E2E optimization for more effective and nuanced image communication. Authors in [69] introduce a DeepJSCC scheme designed primarily for effective image transmission across wireless networks. The model considered the channel as a non-differentiable constrained layer, using an autoencoder (AE) as shown in Fig. 9. Their DNN-based technique outperformed standard image compression methods such as better portable graphics (BPG), especially when paired with sophisticated channel coding like low-density parity-check (LDPC), demonstrating notable effectiveness in low SNR conditions. Subsequent developments and enhancements to the original scheme were introduced in [70]. To address the challenges of fluctuating bandwidth and decentralized source coding in DeepJSCC, innovative strategies for adaptive-bandwidth image transmission [71], [72] and decentralized transmission [73] have been introduced. Moreover, considering the traditional feedback system, [74], [75], [76] propose image transmission with channel output feedback. [74] reveals that incorporating feedback can significantly boost the efficiency of JSCC in traditional communication systems, and the use of variable-length coding can further enhance the transmission performance. Building upon this, [75] applied these principles to a CNN-based DeepJSCC framework, where the incorporation of a feedback system led to even more impressive results. The performance further improved drastically when considering variable-length coding, leading to a 50% reduction in overall bandwidth [75]. However, the study had limitations such as high complexity, inadaptability, suboptimality and non-generalizability, which were addressed in [76] by implementing a singular encoder-decoder at the transmitter and receiver, respectively. Additionally, [76] leverages vision transformer (ViT) as shown in Fig. 10, which utilizes a global self-attention mechanism for more distinct representation of semantic features, surpassing the performance of CNN-based strategies that learn the semantic features of images from a local to global context with a restricted receptive field. The DeepJSCC schemes for image transmission have been practically validated in communication system testbeds in [70], [77]. The authors in [77] introduced a real-time semantic testbed based on a ViT, which was later expanded in [78] with an software-defined radio (SDR)-based wireless SemCom prototype.

Recent advancements have demonstrated the effectiveness of DeepJSCC models, particularly through improvements with channel output feedback [75], [76] and refinement methods [71], [73], [79]. Despite these developments, the models face two significant challenges: SNR-adaptation and code rate (CR)-adaptation problems. The SNR-adaptation problem is that DeepJSCC models, trained at a specific channel SNR, perform best only when the test and training SNRs are the same, necessitating multiple models (and training) to cover the full SNR regime. The CR-adaptation problem is the inability to dynamically adjust the CR of semantic codes in a DeepJSCC model, which is necessary for optimizing bandwidth usage across different channel SNRs and image contents. The two problems necessitate the training of several DeepJSCC models to accommodate all channel SNR and CR regions which require high computation. Moreover, even with a sufficient array of DeepJSCC models trained across a range of channel SNRs and CRs, achieving adaptive rate control for specific image transmission tasks, especially those with quality constraints, remains a challenge due to the unpredictable quality of the reconstructed images. To address the SNR-adaptation problem the authors in [80] developed the attention module-based DeepJSCC (ADJSCC) model, which is based on channel coding rate, compression ratio optimization, and utilization of squeeze-and-excitation scheme to achieve SNR-adaption. ADJSCC optimizes coding resources between source and channel coding/decoding processes, enhancing compression efficiency or transmission reliability based on the SNR region. Simulation results show that the peak signal-to-noise ratio (PSNR) performance of a single ADJSCC model matches the highest PSNR achieved by various JSCC models, each trained at different SNRs. This indicates that the ADJSCC model effectively addresses the SNR adaptation issue. To address the CR-adaption problem various variable-length transmissions are proposed in [81], [82], [83]. The authors in [81] developed DeepJSCC-A for wireless image transmission, which utilizes two separate encoders for source and channel coding processes. The resulting semantic codes are split into selective (can be active or non-active) and non-selective (always active) features for transmission. A policy network decides which selective features to transmit with the non-selective ones. The authors in [82] proposed a model with adaptive-bandwidth transmission across various features based on their entropies. This technique allocates higher bandwidth to key features and improves the quality of image reconstruction. Moreover, the strategy for allocating bandwidth is simultaneously learned with the non-linear transformation in an E2E manner, which further improves the performance of the model significantly. The authors in [83] address SNR-adaptation, CR-adaptation, and the transmission quality prediction problem to achieve transmission quality-guaranteed rate control in every transmission. They propose a predictive and adaptive deep coding (PADC) framework for achieving flexible, adaptive, and quality guaranteed rate control according to different channel SNR and image contents. The PADC scheme achieved superior performance compared to previous techniques.

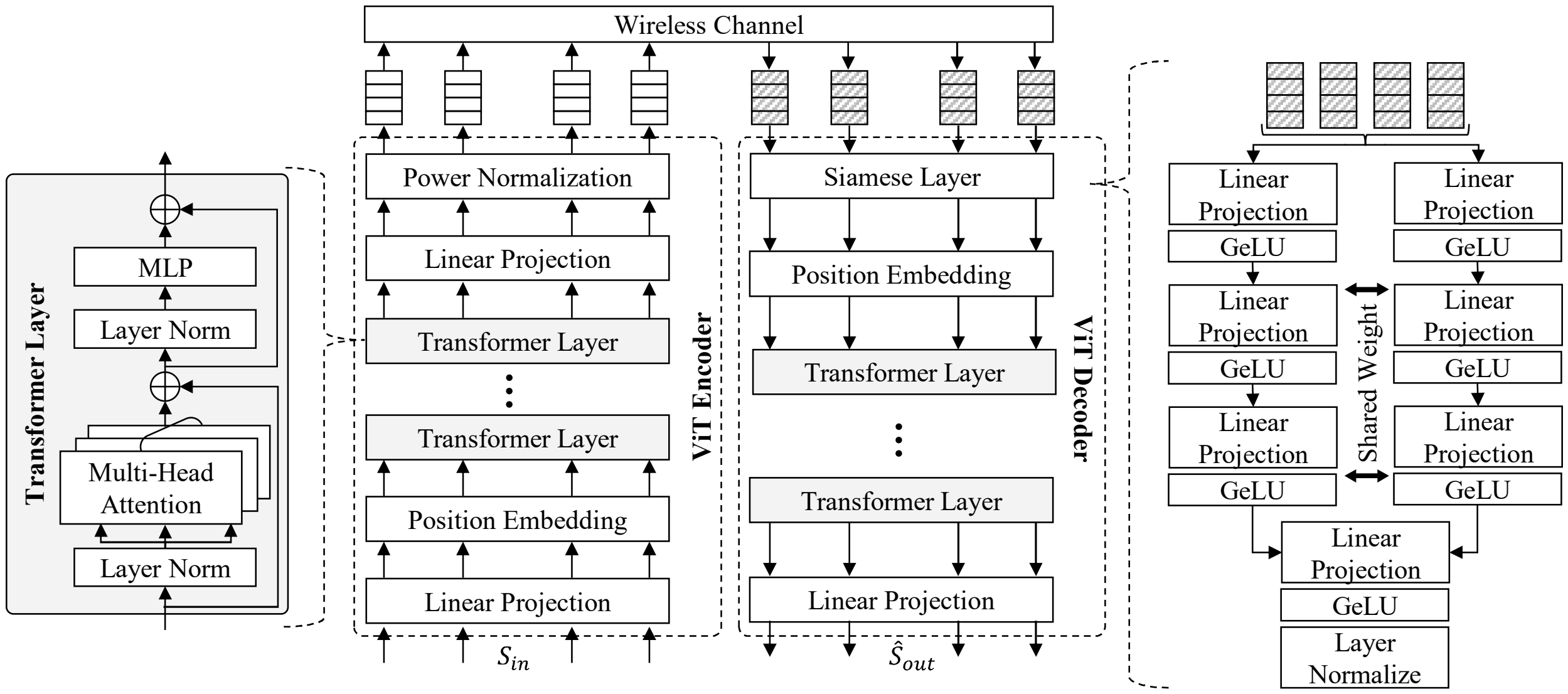

**FIGURE 10. A transformer-based architecture with symmetric structure for encoding and reconstructing the source signal [76].**

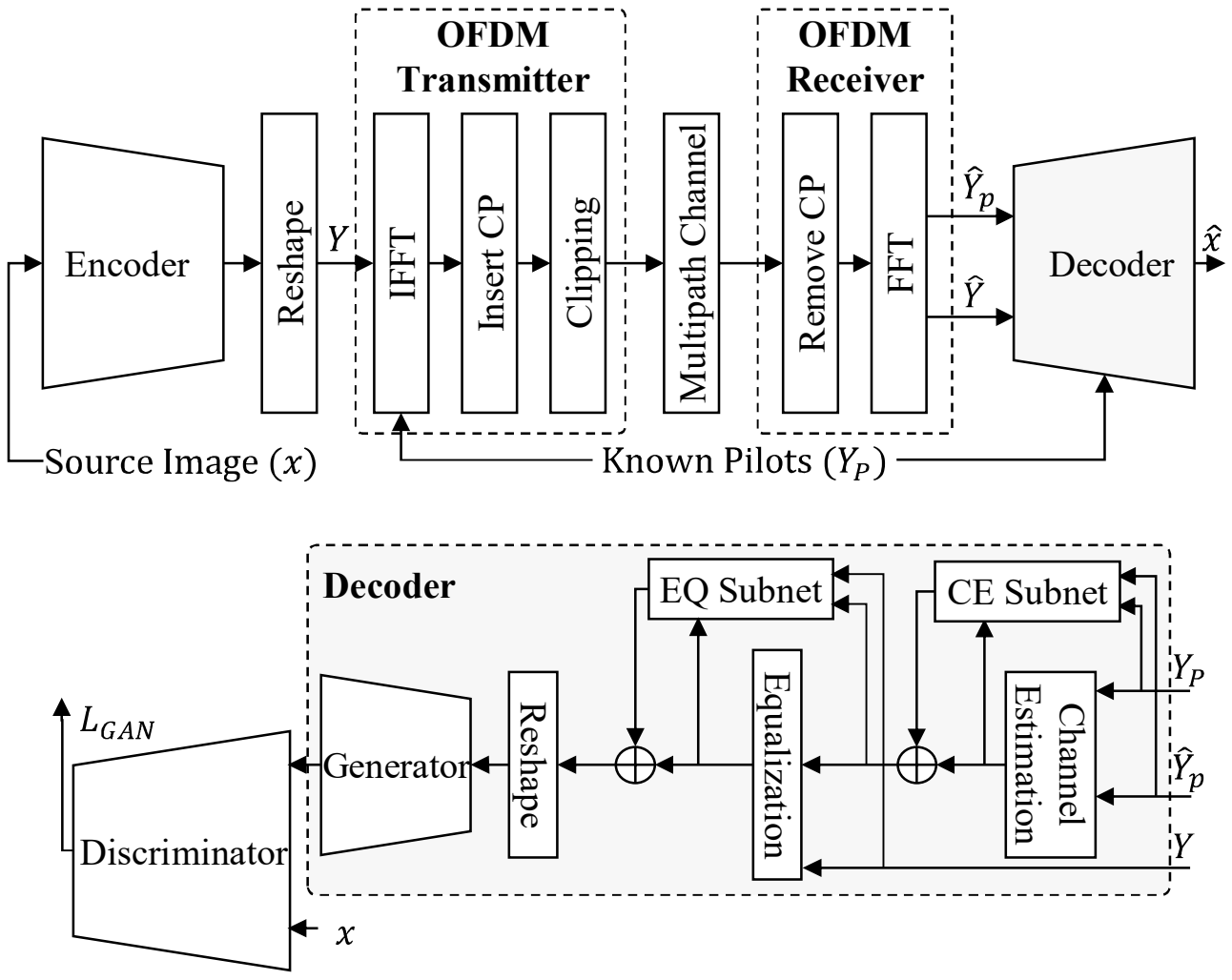

**FIGURE 11. DeepJSCC based OFDM system (Top) and the decoder design (Bottom) with GAN structure. Where, $Y$ is the frequency domain complex-valued symbol, and $\hat{Y}$ and $\hat{Y}_p$ are the frequency domain symbols and pilots, respectively [84].**

DeepJSCC-based SemCom systems for image transmission have been expanded to include orthogonal frequency division multiplexing (OFDM) communication systems [84], [85], where signal processing blocks such as channel estimation and equalization are considered for the domain knowledge as illustrated in Fig. 11. It has also been expanded to multiple-input multiple-output (MIMO) channels [33], [86], [87], [88] with various antennas, and non-orthogonal multiple access (NOMA) system [89]. Authors in [90] introduce a novel framework aimed at boosting resource efficiency in two-user uplink NOMA systems. This framework employs SemCom for a secondary, distant user (F-user), while maintaining traditional bit-based communication for a primary, closer user (N-user). The study examines the semantic-versus-bit rate region, revealing that SemCom significantly enhances the performance of F-user under stringent power constraints, though it may not perform as well with higher power allocations. Key contributions of the research include the development of a dual approach—combining SemCom and bit-level communication—that optimally adjusts to channel conditions, and a thorough performance comparison with traditional NOMA frameworks. The authors show that semantic-enhanced NOMA systems greatly improve communication efficiency and overall system performance, particularly when the primary user faces power limitations. This highlights the benefits and challenges of integrating SemCom into NOMA networks. When extending to OFDM schemes, although most researchers permit encoders to map input signals freely without adhering to any specific input constellation, real-world digital communication systems usually employ a fixed constellation design for symbol mapping. Implementing a full-resolution constellation design for modulation in radio frequency (RF) systems presents significant challenges. The authors in [91], [92], [93] address the challenges of constellation design for image transmission in DeepJSCC-based SemCom.

Authors in [94] propose a multi-task scheme that simultaneously recovers and classifies images under various channel conditions, using a DL approach that optimizes both tasks through JSCC. By incorporating a novel training scheme that minimizes MSE and reduces CR, their system demonstrates improved resilience against channel variability, leading to enhanced performance in image recovery and classification tasks compared to single-task systems. This method exhibits enhanced resilience against channel noise and shows improved performance in image classification accuracy and recovery quality under different SNR levels. However, the reliance of the system on specific image data types and optimized channel conditions may limit its generalizability and applicability in diverse real-world scenarios. This is particularly true for devices with limited computational power, thus restricting its broader deployment in edge computing and IoT environments.

For practical implementation of SemCom systems, authors in [95] proposed a deep separate source-channel coding method tailored for SemCom tasks. This framework utilizes a variational AE to address rate-distortion problems while considering semantic aspects, aiming to optimize both data recovery and classification performance. The main contribution is the development of a Bayesian model-based rate-distortion optimization framework that integrates semantic tasks with general data distributions. For practical uses like image transmission and classification, the authors have introduced a forward adaptation scheme combined with variational autoencoding. This enhancement improves feature extraction and better adapts to the density of image features. Simulation results demonstrate that this scheme outperforms traditional compression methods and other DL-based approaches. However, the study recognizes limitations, including a dependence on extensive training data for optimizing DL models and challenges in generalizing across various semantic tasks and data distributions.

The aforementioned image SemCom applications with JSCC schemes mainly focus on either data reconstruction or semantic coding aspects. However, they do not emphasize the semantic similarities between the original signal and its reconstruction at the receiving end. Instead, they primarily regard the MSE or the structured similarity metrics as the E2E distortion measure. Another aspect of SemCom is task execution, where the receiver is not necessarily interested in reconstructing the original source signal with minimal distortion, but rather is focused on specific downstream tasks or applications. However, impairments from different PHY in communication can heavily corrupt the received signal and make the system incompatible for the downstream tasks [41]. E2E learning in task-oriented SemCom enables task execution despite PHY impairments, without the need to construct the original source. For instance, the image classification-oriented SemCom system described in [96] enhances recognition accuracy; similarly, the system in [97] aimed at image reidentification of a person improves retrieval accuracy. Additionally, the SemCom approach in [98] is tailored for edge inference tasks related to classification, while the system presented in [99] is specifically designed for multimodal visual question answering tasks. In contrast, in contexts where extreme image compression is necessary [100], the receiver may focus more on producing an output that mirrors the original signal distribution, instead of striving for an exact replication of the source for the task at hand. This concept, referred to as perception loss in studies of image compression, plays a crucial role in navigating the trade-off among rate, distortion, and perception [101].

Generally, perception loss is evaluated by the difference between the original and reconstructed distributions, which can be reduced using GANs [102]. The primary objective of classical DeepJSCC is to accurately recreate the original signal at the receiver, minimizing distortion by optimizing both source and channel coding simultaneously with DNNs. However, this approach can result in reduced perceptual quality, adversely affecting performance in subsequent downstream tasks. To solve the problem, the authors in [103] implemented a context-aware communication strategy, utilizing StyleGAN generators and distortion metrics that closely mimic human perception throughout the optimization process. Building on their approach, initially, the authors introduced InverseJSCC. This method uses a pre-trained GAN generator to tackle the wireless communication challenge from a new angle, enhancing perceptual quality through unsupervised learning. Following this, they developed GenerativeJSCC, offering a supervised E2E solution for wireless image communication by leveraging a pre-trained GAN, further refining the strategy to address the outlined problems. Their simulation results indicate that the proposed methods outperform the DeepJSCC model for the downstream tasks in terms of both distortion and perceptual quality.

### *E. VIDEO APPLICATIONS*

JSCC for video transmission, first proposed in [104], outperformed separate coding schemes in practical multimedia applications. The authors highlight the complexity of delivering quality of service (QoS) for video communication applications, particularly considering packet loss and data corruption due to network congestion or physical channel impairments. To address these issues, the paper advances the idea of JSCC, where redundancy is introduced in both source and channel coding to enhance error-free, real-time video communication systems. Furthermore, the authors explore cross-layer design that operates adaptively to changing network conditions, and proposed a power-minimized bit-allocation scheme that processes power for source coding, channel coding, and transmission with the aim of optimizing overall system performance.

The authors in [105] were the first to apply a DeepJSCC framework for video transmission to enable SemCom, naming it DeepWiVe. The transmitter directly maps video signals to symbols and combines video compression, channel coding, and modulation blocks as a DNN for an E2E optimization. The decoder at the receiver predicts residuals without any distortion feedback by considering occlusion/disocclusion and camera movements, which enhances the video quality. Additionally, the system simultaneously trains on different bandwidth allocation networks for the video frames to facilitate various bandwidth transmissions, and later trains the bandwidth network using RL for optimizing bandwidth allocation. DeepWiVe significantly improves video and visual quality and outperforms both uniform allocation and heuristic policies by a wide margin. Experimental results demonstrate that DeepWiVe effectively mitigates the cliff-effect, a common limitation in traditional digital communication, ensuring smoother performance degradation under varying channel conditions. Moreover, in motion detection tasks, DeepWiVe surpasses conventional compression protocols like H.264+LDPC and H.265+LDPC, improving the multi-scale structural similarity index measure (MS-SSIM) scores by margins of up to 0.0485 and 0.0069, respectively. However, the complexity of its neural network architecture requires extensive computational resources and training, which could limit its application in real-time or resource-constrained environments. Moreover, despite its advanced prediction capabilities, there is an implicit limitation in scenarios with highly variable or unknown channel conditions, where the absence of explicit error correction could lead to performance degradation. This reliance on large datasets for training also raises concerns about data availability and privacy, potentially restricting its usability in sensitive applications.

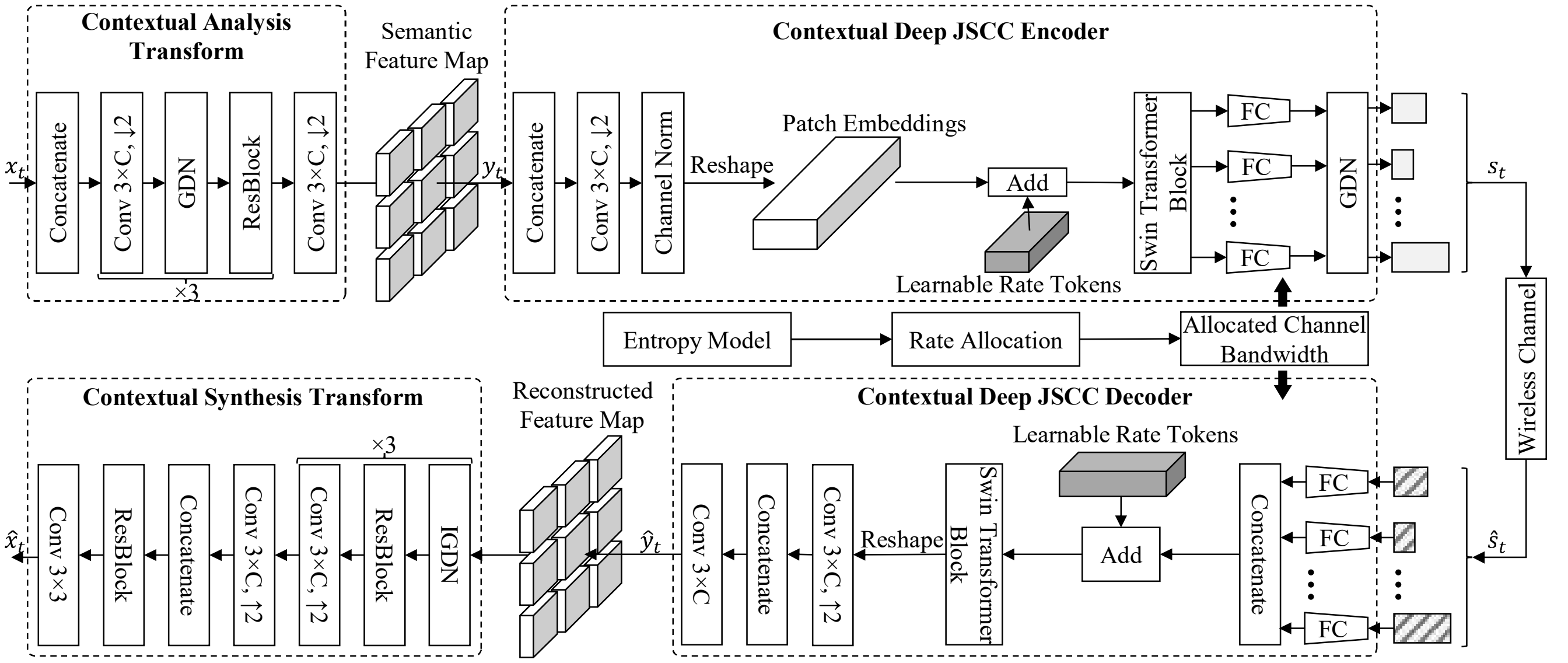

**FIGURE 12. Semantic video transmission with E2E optimization. Where, ↑ 2/↓ 2 is upscaling or downscaling with 2 strides. $x_t$ are input frame, $\hat{x}_t$ are output frame, $y_t$ are transmitted semantic feature, $\hat{y}_t$ are received semantic feature, $s_t$ are channel input vector, and $\hat{s}_t$ are received sequence, all at time step $t$. FC is Fully Connected layer, GDN is Generalized Divisive Normalization and IGDN is Inverse GDN [106].**

The authors in [106] proposed deep video semantic transmission (DVST) method to further improve the efficiency and robustness of E2E video transmission framework using DeepJSCC as shown in Fig. 12. DVST employs a nonlinear transform [107], converting high-dimensional sources, like video, into latent representations. This process, facilitated by variational latent-variable models, is designed to extract semantic features from across video frames. Subsequently, DVST transmits these extracted semantic features through wireless channels utilizing DeepJSCC. Integral to this method is the incorporation of a temporally adaptive entropy model. This model orchestrates the transmission of the current frame, leading to a more robust and accurate entropy model compared to earlier versions. Moreover, it utilizes a rate-adaptive transmission mechanism, efficiently allocating the limited channel bandwidth among video frames to maximize overall transmission performance. Experimental results show that DVST saves up to 50% in channel bandwidth costs compared to traditional schemes such as H.264/H.265 + LDPC and digital modulation. DVST achieves superior results in segmentation and reconstruction across a range of channel bandwidth ratios. Its performance was confirmed through downstream computer vision tasks, demonstrating high fidelity results applicable to both computer vision and human vision. Nevertheless, dependency of DVST on the accurate extraction and interpretation of semantic features may limit its effectiveness in complex visual scenarios where the semantic information is not distinctly defined or is prone to misinterpretation. This could potentially affect the accuracy and reliability of the transmitted video content, with a trade-off between efficiency and the robustness of transmission quality under such conditions.

When compared to traditional methodologies, both DeepWiVe and DVST offer substantial advancements in handling network variations and optimizing bandwidth. These methods significantly outperform standards like H.264 and H.265 in terms of structural similarity indices, showcasing their potential to improve visual quality while reducing bandwidth requirements. However, the computational efficiency of these methods, while improving, still lags behind some traditional methods, particularly in environments that demand low latency and high-speed processing. Both frameworks also introduce complexities in model training and require robust hardware support, which could limit their applicability in resource constrained environments.

A different approach for semantic video application is proposed by the authors in [108]. They utilize driving videos, employing DL techniques to extract features from these videos and integrate them into images for the purpose of animation and video synthesis. The *keypoints* that correspond to the expression or motion are extracted from the frames of a driving video to be transmitted, and a generator uses these keypoints to animate the source image into those expressions or motions. The authors in [109] propose a DL model to synthesize a talking head from a facial driving video and extend its application to video conferencing. In this context, the keypoints transmitted from the facial driving video are utilized to recreate the driving video by combining these keypoints with an image of the speaker, effectively synthesizing the video content. Although these studies consider semantic extraction for video synthesis using a DL-based source coding module, they do not account for a communication system for keypoint transmission or errors due to varying channel conditions in semantic video conferencing (SVC). Therefore, the authors in [110] consider a communication system for transmitting keypoints in SVC, and Fig. 13 illustrates the three-level architecture of the SVC. SVC drastically reduces transmission bandwidth by transmitting only keypoints instead of the entire data stream; thus, any error in transmission will lead to a change in expression, as opposed to pixel distortion seen in conventional systems. To detect the degree of expression and changes due to transmission noise in the keypoints, the authors developed an incremental redundancy HARQ framework for SVC (SVC-HARQ), tailored for varying channels to identify semantic errors. Additionally, they exploit CSI to automatically allocate varying levels of information across different subchannels, training the system to allocate more information to subchannels with higher SNR and, less allocation to those with lower SNR. The experimental results showed that the system achieved higher resource efficiency and video quality compared to existing methods, along with robustness against varying channel conditions. While the system demonstrated superior video quality and resource efficiency, its reliance on accurate CSI for optimal performance may limit its effectiveness in environments where channel conditions fluctuate rapidly or are unpredictably noisy. Further refinement could focus on enhancing the robustness of the HARQ mechanism to maintain performance even when CSI is imperfect or outdated.

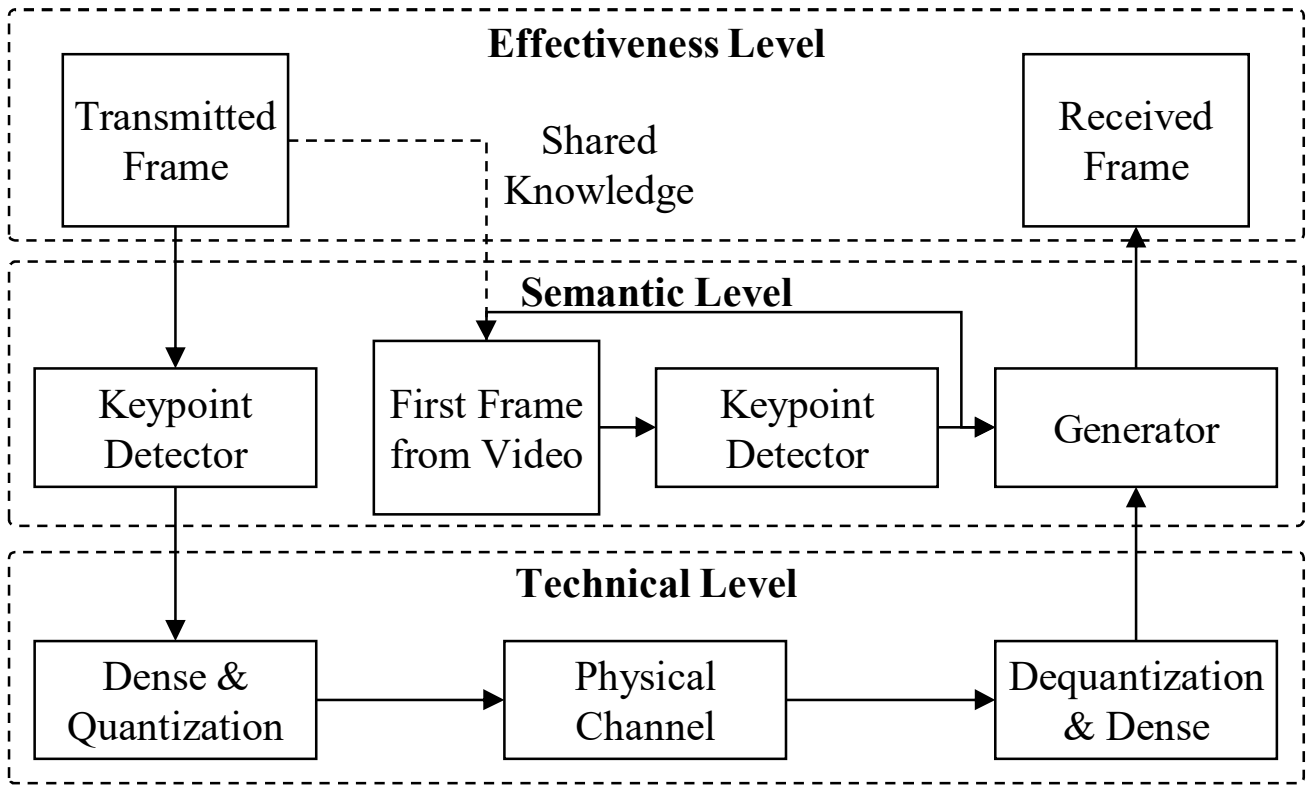

**FIGURE 13. SVC in three-level architecture [110].**

Point cloud video (PCV) streaming is an emerging application with 6-degree of freedom (DoF) and higher data volume, compared to other streaming forms such as 2-dimensional video (0-DoF), 360-degree video (2-DoF), or virtual reality video (3-DoF). However, it requires a massive amount of raw PCV frames and high bandwidth capacity. Even after employing traditional compression techniques (lossy, lossless, or transmission tiles), the bandwidth demand reaches gigabytes per second, exceeding the capacity of current 5G networks. Additionally, the decoding of this volumetric media demands substantial computational overhead. DL techniques can be applied in PCV streaming to identify and transmit only key features to the endpoints, where they are reconstructed through a GAN-based generative network. This process effectively utilizes SemCom to enhance efficiency and reduce bandwidth requirements. The authors in [111] proposed AITransfer scheme to enable semantic-aware transmission for the real-time PCV services. AITransfer integrates the bandwidth of dynamic networks into the design of an E2E architecture for SemCom and deploys an online adapter to optimize the compression ratio of the inference model. This approach adapts to the network conditions with 30 times improvements compared to conventional techniques. Despite its effectiveness, the requirement for consistent high-bandwidth conditions by AITransfer could limit its practical deployment in environments with fluctuating network capabilities. Further advancements could investigate optimizing the adaptability of the system to varying network strengths without compromising transmission quality.

While AITransfer achieves superior video quality output, its resource-intensive nature renders it unsuitable for devices with limited resources. The authors in [112] introduced a new SemCom framework named ISCom to enhance PVC streaming on resource-constrained mobile devices. ISCom framework incorporates an interest-aware SemCom scheme, two-stage region of interest (ROI) selection methods, a lightweight deep encoder-decoder trained with AITransfer as a pre-trained model, and an artificial intelligence (AI) codec scheduling method. The study demonstrates that these methods enhance streaming efficiency, reduce energy consumption in mobile devices, and enable the encoder-decoder network to adapt to diverse network environments through a deep RL-based intelligent scheduler. Experimental results indicate an improvement ranging from a minimum of 10 frame per second (FPS) to a maximum of 22 FPS over existing methods on resource-constrained devices. However, the computational demands of the AI-driven scheduling component and the initial setup requirements for training the deep RL model could limit rapid deployment or scalability. Future developments might focus on simplifying the training phase and optimizing the scheduler to operate with minimal computational overhead while still achieving high efficiency.

The authors in [113] explored the potential and challenges of real-time holographic video communications, essential for immersive experiences in the future metaverse. The authors review state-of-the-art holographic PCV transmission techniques and highlight the limitations of current 5G networks for PCV transmission. They propose an AI-driven solution that involves a novel transmission mechanism, E2E network design integrating encoding and decoding, and adaptive streaming technology. The proposed method achieves a compression ratio of up to 30 times, surpassing traditional methods. Across different bandwidth, the AI-driven solution achieves lower transmission latency while maintaining higher quality of experience (QoE). The AI-driven method also maintains high FPS across varying computational capabilities, significantly outperforming the benchmark. Additionally, the dynamic reinforcement learning-based online controller ensures optimal model selection in real-time, adapting to network conditions to sustain high QoE. However, limitations remain in real-time processing capabilities on mobile devices and the extensive computational resources required for training AI models. Future research directions suggested interest-aware PCV capturing, extending AI-driven transmission across other standards, and developing comprehensive quality assessment methods for AI-driven wireless transmission.

### F. AUDIO APPLICATIONS

The authors in [114] introduced a novel DL-based SemCom for speech signals, named DeepSC-S. The key feature of the system is its focus on minimizing semantic errors over bit or symbol errors, significantly enhancing transmission efficiency. DeepSC-S employs an attention mechanism within a squeeze-and-excitation network, assigning higher weights to semantic information (key speech information) during training, which leads to improved signal recovery. Furthermore, a versatile DL model was developed, ensuring effective operation of DeepSC-S across a range of channel conditions, particularly in areas of low SNR. The robustness and adaptability of DeepSC-S were validated through tests using both telephone and multimedia transmission systems. In the telephone system, DeepSC-S achieves 10.92% and 7.34% better performance in SDR and perceptual evaluation of speech quality (PESQ), respectively, compared to benchmark systems. A similar trend was observed in the multimedia transmission system, however, there were more overlapping points under the AWGN channel. Overall, DeepSC-S surpassed benchmark performances at lower SNRs across all channels in both telephone and multimedia systems. However, at higher SNRs under the AWGN channel, other benchmarks showed better performance in some cases. The limitations of DeepSC-S include its potential underperformance in high

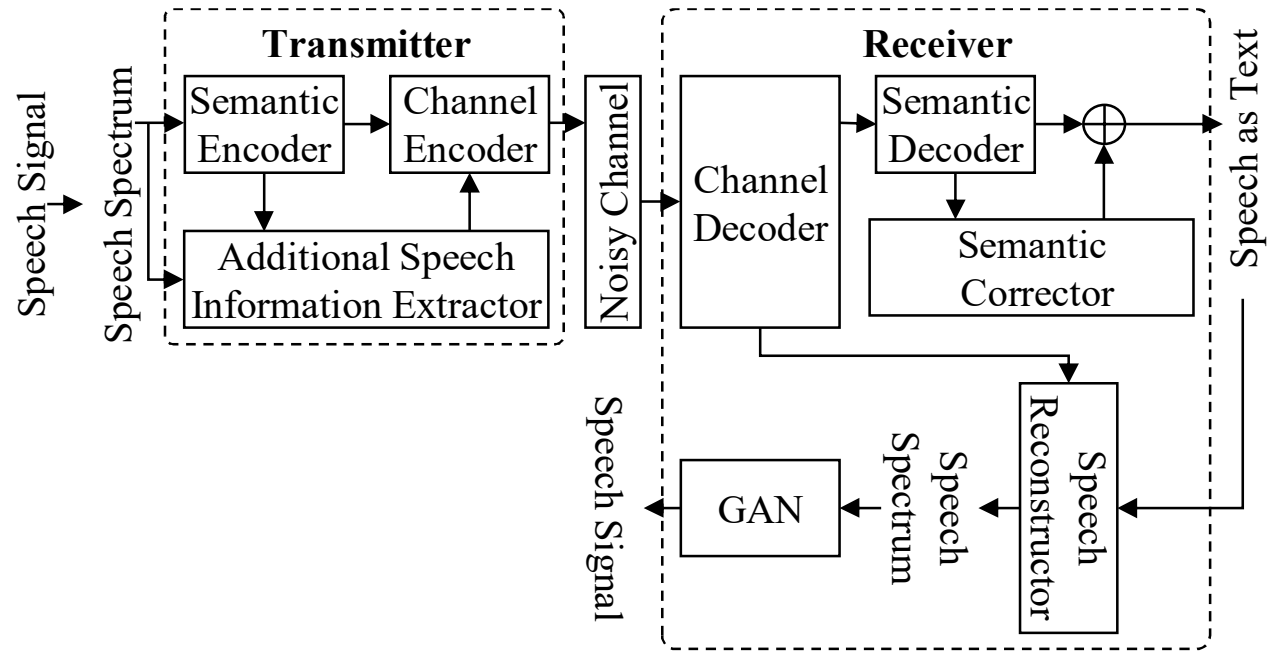

FIGURE 14. **Speech transmission with E2E optimization [116].**

SNR environments where traditional systems might still have an edge in accuracy. Further refinement of the model could focus on enhancing performance across a broader range of SNR levels to uniformly outperform traditional benchmarks.

In [115], the authors expanded their research to include DL-based SemCom for speech transmission, introducing DeepSC-ST. In this system, a joint semantic-channel encoder, leveraging CNN and RNN-based semantic transmitters, extracts semantic features as text from the speech input. This approach reduces the data needed for transmission without compromising system performance. At the receiver end, the semantic information is retrieved from the text and transformed into full text using a feature decoder. Using the full text and speaker information, a CNN and RNN-based model performs speech synthesis to regenerate the speech signals. Experimental results showed that DeepSC-ST outperforms all other benchmarks, showing significant improvement in the range of -8dB to 8dB across all channel models. Additionally, the system is robust and adaptable to dynamic channel conditions in low SNR regions. However, the reliance on complex DL models may introduce challenges in real-time processing and require substantial computational resources, potentially limiting deployment in low-resource environments. Optimizing the model for faster processing with fewer resources could further enhance its practicality.

The authors in [116] proposed a DL-based E2E communication system for speech transmission focusing on transmitting both semantic-relevant and semantic-irrelevant information for efficient speech recognition and speech reconstruction tasks, respectively. As illustrated in Fig. 14, the system used an attention-based soft alignment module to extract solely text-related semantic features, and a redundancy removal module to omit redundant semantic features. The system also employed a pre-trained language model as a semantic correction module to accurately predict transcription with the semantic information, thereby further improving accuracy. For speech-to-speech transmission, the system included a connectionist temporal classification (CTC) alignment module to extract additional speech-related but semantically irrelevant information, such as pitch, time intervals, power, and speaker identification, to reconstruct the original speech signals at the receiver. Additionally, the authors introduced a two-stage training scheme to accelerate the training of the DL model. Experimental results showcased the remarkable transmission efficiency of the proposed scheme, with only 16% and 0.2% of the total symbols required for speech-to-text and speech-to-speech transmissions, respectively, while still maintaining signal quality. A potential limitation is the dependence of the system on extensive training data to achieve high accuracy and robustness. Enhancing the ability of the system to perform well with limited training data or under more varied acoustic conditions could be a valuable area for future research.

In study [117], the authors explored the application of audio SemCom across extensive wireless networks, specifically emphasizing edge devices transmitting large-scale audio data to a server. They proposed a SemCom model utilizing a wave-to-vector (wav2vec) based AE architecture comprising CNNs. Designed to conserve bandwidth, this model transmits only the contextual features of the audio signal as semantic information by integrating source and channel coding with an AE. At the transmitter, the temporal audio signal feature is encoded into semantic information for transmission over an imperfect, noisy wireless channel. At the receiver, the decoder recovers the audio signal from the semantic information, effectively mitigating noise interference. Additionally, the authors use FL to improve the accuracy of semantic information extraction. In FL, each local model is trained with audio data from its corresponding device, and the parameters of these models are sent to the server to be amalgamated into a global model. This method integrates audio features from multiple users, significantly improving the precision of semantic information extraction. Experimental results highlighted the effective convergence of the algorithm and its ability to reduce the MSE of audio transmission by nearly 100 times compared to traditional coding schemes. However, the study acknowledged certain limitations such as potential scalability concerns, dependency on the quality and variability of the audio data from each local device, and the robustness of the system in real-world noisy conditions. Future efforts could focus on refining the FL to enhance scalability and robustness, ensuring consistent performance across a wide range of network conditions and participant variability.

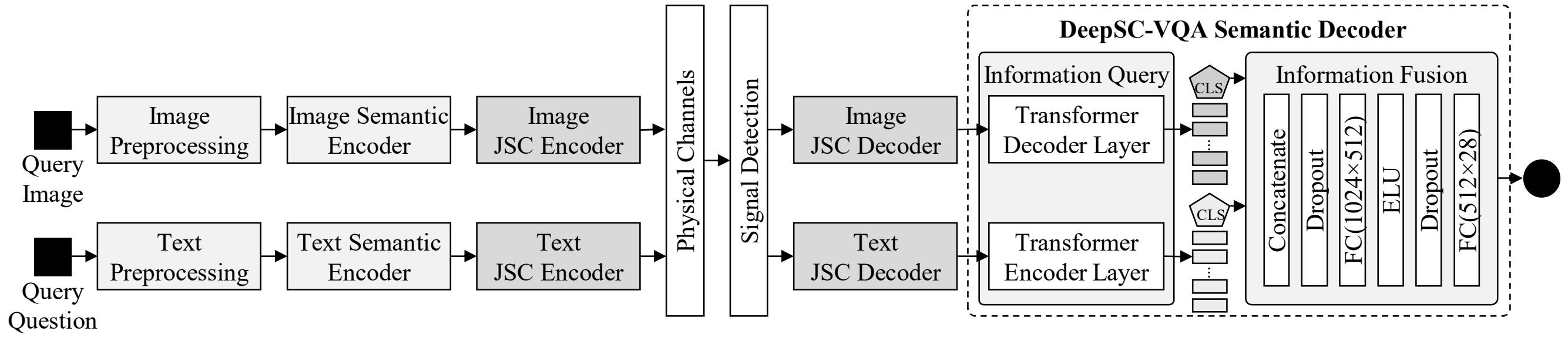

**FIGURE 15.** **Multi-user multi-modal SemCom structure using DeepSC-VQA transceiver [99].**

### *G. MULTIMODAL APPLICATIONS*

In multimodal applications, the system model incorporates various data types, including text, image, video and audio. In [118], the authors presented MUDeepSC, a multi-user, task-oriented SemCom system for visual question answering (VQA), specifically designed to utilize multimodal data from multiple users. Compared to traditional communication systems, this system demonstrates significantly greater resilience to channel variations, particularly in low SNR scenarios. Building on this concept, the authors in [99] developed a multi-user SemCom system that supports both single-modal tasks, such as image retrieval and machine translation, and multimodal tasks like VQA. They designed three transformer-based transceivers — DeepSC-IR, DeepSC-MT, and DeepSC-VQA — all utilizing the same transmission architecture but differing in their receiver designs. The structure of the DeepSC-VQA system is depicted in Fig. 15. When trained using specific algorithms, these transceivers surpassed traditional ones, particularly in low SNR environments. Their SemCom system features a multi-user MIMO setup, with multiple antennas and transmitters, each utilizing a DL-based semantic encoder. Depending on the type of receiver, this system can manage either single-modal or multimodal data transmission for multiple users. Within this framework, data from various users complement each other semantically. The systems employ a linear minimum mean squared error (L-MMSE) detector for signal retrieval and a joint source-channel (JSC) decoder to minimize channel distortion and reduce inter-user interference. This design facilitates the joint execution of tasks such as image retrieval and machine translation, with multimodal semantic receivers integrating data from multiple users.

The study [119] explores SemCom techniques designed for multimodal signal transmission, focusing on cross-modal SemCom, which is essential for efficiently applying semantics to text, audio, image, and video signals. For 6G applications that prioritize immersive experiences, transmitting multimodal signals with minimal latency and increased reliability is crucial. As outlined in [119], cross-modal SemCom incorporates a cross-modal semantic encoder, a decoder, and a cross-modal knowledge graph (CKG) to achieve this goal. The encoder processes video, audio, and haptic inputs to generate both explicit and implicit semantics for transmission. The decoder maintains signal consistency at both bit and semantic levels. The CKG serves as a knowledge base (KB) that supports both encoding and decoding processes. Additionally, the system utilizes GAN-based models to recover signals. However, these models occasionally produce incomplete outputs due to various distortions. Consequently, the authors propose a signal completion module to improve the quality of recovered multimodal signals by retrieving similar patches from the CKG. However, a significant challenge is the potential semantic ambiguity of the recovered signals. To address this, authors of [119] suggest optimization approach using RL that utilizes both semantic and bit similarity as rewards. This methodology, demonstrated through the deep Q-network (DQN) algorithm, combined semantic and bit similarity leading to more accurate signal recovery in SemCom.

The authors in [120] present an advancement in multimodal, multi-user, goal-oriented SemCom by introducing a cooperative system designed for Internet of Vehicle (IoV) applications, such as pedestrian detection, traffic analysis, and vehicle tracking. The architecture includes several key components – a semantic encoder that extracts meaningful data, a cooperative semantic decoder that reconstructs original data based on specific goals and user interactions, a JSC encoder that converts semantic data into channel input symbols, a cooperative JSC decoder designed to retrieve semantic information from multiple users, and a semantic-driven cooperative task performer that accomplishes specific tasks by adapting to the semantic information received from various users, utilizing semantic correlations and unique user features. This SemCom system incorporates pre-learned user correlations and requires sharing knowledge between transmitters and receivers. Importantly, each user possesses a background KB that is shared via joint training on a communal dataset.

Within the context of emerging 6G applications such as augmented reality/virtual reality (AR/VR) and online role-playing games, the authors in [121] introduce a multi-access edge computing (MEC) framework designed for goal-oriented multimodal SemCom. This framework integrates a bidirectional caching task model that is well-suited for DL applications. The authors propose a caching-enhanced offloading method to lower computational costs, formulating it as a mixed-integer non-linear programming problem. Subsequently, they present the content popularity-based DQN (CP-DQN) caching algorithm for near-optimal caching decisions, along with the cache – computation coordination optimization algorithm (CCCA) to strike a balance between computing and caching. These algorithms have been proven optimal in terms of cache hit rate, rewards, and overall cost.

The transmission of three-dimensional (3D) contents to provide immersive communication experiences became increasingly important for emerging 6G applications such as the AR/VR. The authors in [122] developed a novel generative AI model assisted 3D semantic communication (GAM-3DSC) system for 3D SemCom in 6G networks. The key contributions include the development of a 3D semantic extractor (3DSE) that utilizes advanced AI models like the segment anything model (SAM) and neural radiance field (NeRF) to extract essential 3D semantics from user-defined perspectives. Additionally, an adaptive semantic compression model (ASCM) is introduced to encode and reduce redundant semantic information, while a conditional generative adversarial network and diffusion model aided channel estimation (GDCE) is designed to refine CSI for better signal recovery. The system demonstrates significant improvements in transmitting goal-oriented 3D scenarios. GAM-3DSC demonstrates notable experimental results, extracting key 3D semantics and reducing data transmission volume by up to 80% with ASCM while maintaining high image quality with PSNR around 25 dB and structural similarity index measure (SSIM) of 0.95. It also outperforms traditional and DL methods with GDCE, achieving the lowest normalized mean squared error (NMSE) across all SNRs. Throughout the experiment, It ensured high semantic consistency between transmitted and received 3D scenarios with a bilingual evaluation understudy (BLEU) score of 0.61 and cosine similarity of 0.97. However, limitations include potential computational complexity and the need for extensive training data for the generative models to function optimally. Prior to GAM-3DSC, the authors proposed large AI model-based semantic communication (LAM-SC) framework designed specifically for image data in [123]. LAM-SC addresses the challenges of KB construction including limited knowledge representation. The key contributions include the development of SAM-based KB (SKB) to split images into semantic segments, an attention-based semantic integration (ASI) mechanism to weigh and integrate these segments, and an adaptive semantic compression (ASC) encoding to reduce communication overhead by eliminating redundant information. The findings from simulations demonstrate that LAM-SC significantly reduces data size while maintaining high image quality in terms of PSNR and SSIM values compared to traditional methods.

The authors in [124] provided a comprehensive overview of integrating NeRF and 3D Gaussian splatting (3D-GS) to enhance immersive communication through photorealistic 3D content rendering. It discusses over-the-air training using FL within a device-edge-cloud architecture and outlines practical rendering architectures. They also propose a novel SemCom-enabled 3D content transmission design and demonstrate the application of radiance fields in wireless environments, such as radio mapping and radar imaging. The proposed framework outperformed benchmark schemes in average PSNR at higher communication rates with feature selection and prediction. The framework also improved rendering efficiency and model compression. However, there are few limitations in the proposed framework such as similar PSNR score in higher and lower communication rate regimes, high computational demands for system operations, and challenges in communication and resource management within distributed wireless network.

Development of multi-task services in SemCom systems is crucial for modern data-driven and downstream AI applications. The integration of multimodal, multi-user, and multi-task approaches in SemCom systems, as previously discussed, significantly enhances the efficiency and adaptability of these systems in complex network environments. Building upon this foundation, authors in [125] propose a comprehensive approach to enhance multi-task capabilities in SemCom systems, particularly focusing on the simultaneous execution of data recovery and AI tasks. The authors introduce a unified framework designed to handle multiple tasks and data modalities (image, text, and speech) using a single DL model. This framework utilizes a task-aware, DL-based semantic encoder that optimizes data representation for multiple tasks simultaneously. It adjusts the number of transmitted symbols based on task requirements and channel conditions, effectively reducing transmission overhead and model size while maintaining performance comparable to task-specific models in terms of accuracy and data integrity. This approach promises significant improvements in resource management within SemCom networks, particularly in environments with limited bandwidth and computational resources. However, the model complexity and the potential difficulty in adapting to new or different tasks without extensive retraining could limit its practical deployment in real-world scenarios.

Table IV summarizes the key contributions and limitations of SemCom across different modalities. Additionally, it enumerates the channel models, metrics, DL algorithms, and datasets utilized in these systems. To assess the performance of semantic applications, a range of metrics is utilized. For text and speech semantic evaluations, commonly used metrics include the word error rate (WER), bilingual evaluation understudy (BLEU), and character error rate (CER). In contrast, the PSNR, Structural SSIM, learned perceptual image patch similarity (LPIPS), and mean intersection over union (mIoU) are frequently applied for assessing the quality of images and videos Moreover, a wide array of metrics is employed for interpreting semantic information and evaluating SemCom. The authors of [126] provide a comprehensive survey on metrics used for semantic and goal-oriented communication systems. The authors also discuss developments of SemCom across multiple domains.

**TABLE 4. Summary of semantic communication systems across different modalities**

| | Ref. | Key Ideas and Findings | Major Limitations |
|---|---|---|---|
| Text SemCom | [58] | • First DeepJSCC to improve text data transmission, and achieve better WER than baseline<br>• Preserved semantic information even with severe bit restrictions<br>***Ch:** Erasure Channel; **Mt:** WER; **DL:** RNN, LSTM; **Dt:** European Parliament (EP) [127]* | • Inefficient resource management, uses fixed bit length to encode sentences with varying lengths |
| | [59] | • Proposed DeepSC for semantic, E2E text transmission<br>• Enhanced by transfer learning, adapts to various communication scenarios<br>• Outperformed baseline with low runtime, particularly in low SNR regions<br>***Ch:** AWGN, Rayleigh, Erasure, Rician; **Mt:** BLEU, Loss; **DL:** Transformer; **Dt:** EP* | • Assumed infinite constellations<br>• Model updating, and broadcasting problem<br>• Baseline outperform DeepSC above SNR 12dB over AWGN channel |
| | [60] | • Lightweight DeepSC for efficient, semantic text transmission in resource constrained devices<br>• Addressed first limitations in [59] and improved performance in low SNR and fading channels<br>• Achieved 40-times compression ratio, and adaptive to power and latency constraints<br>***Ch:** AWGN, Rayleigh, Rician; **Mt:** BLEU, SER; **DL:** Universal Transformer; **Dt:** EP* | • Low performance compared to [59] as lightweight<br>• BLEU score of constellation size under Rayleigh or Rician channel not discussed |
| | [61] | • Proposed SemCom system for text using adaptive Universal Transformer<br>• Improved flexibility and adaptation to various channel conditions<br>• Outperformed conventional transformer based SemCom schemes<br>***Ch:** AWGN, Rayleigh; **Mt:** BLEU, SER; **DL:** Universal Transformer; **Dt:** EP* | • The system improvement becomes marginal when the SNR exceeds a certain value, resulting in a sustained symbol error rate (SER) floor<br>• Baseline outperform beyond certain SNR |
| | [62] | • Optimized communication by focusing on semantic similarity using reinforcement learning<br>• Demonstrated robustness in large-scale, and noisy environments<br>• Addressed non-differentiability in both channels and communication objectives<br>***Ch:** AWGN, Phase Invariant Fading Channel; **Mt:** BLEU, WER, BERT-SIM, CIDEr, Reward, **DL:** RL, **Dt:** EP, MNIST [128]* | • RL-based system increases training complexity and instability risks<br>• Suitable for centralized, server-level deployments and not suitable for resource-limited communication scenarios |
| | [63] | • Semantics-aware communication framework with semantic encoding and communication<br>• Established JSNC solution and RL-based SemCom mechanism<br>• Achieved lower WER compared to previous transformer based schemes<br>***Ch:** AWGN Channel and Phase Invariant Fading Channel; **Mt:** WER, Distillation Times, Loss; **DL:** RL, Transformer; **Dt:** EP, CIFAR-10 (C-10) [129]* | • Computational complexity and need trade-off between computation – performance |
| | [64] | • Combined semantic coding with HARQ, improving semantic transmission reliability<br>• Used variable-length coding and introduced various schemes (SCHARQ and Sim32) to enhancing performance and error detection<br>***Ch:** AWGN, Rayleigh; **Mt:** SER, WER, Throughput, BLEU; **DL:** Transformer; **Dt:** EP* | • Often receives corrupted semantic information<br>• Higher bits per sentence compared to traditional scheme due to variable length sentences |
| | [65] | • Proposed a SemCom framework and metrics for textual data transmission<br>• Developed algorithms (APPO) to optimize resource allocation and semantic information<br>***Ch:** Rayleigh; **Mt:** MSS, Users, Tokens; **DL:** RL; **Dt:** AGENDA [130], DocRED [131]* | • Did not consider other metrics to verify the results, only considered the proposed one |
| | [66] | • Proposed SemCom strategies to enhance encoding and decoding processes<br>• Strategies ensure improved semantic accuracy and reduced transmission bits<br>***Ch:** AWGN, Rayleigh; **Mt:** BLEU, METEOR, SS; **DL:** CBOW, LSTM; **Dt:** Brown Corpus [132]* | • Not fully optimized as it did not integrate channel coding strategies into the framework |
| | [67] | • Unified framework enhancing convergence rate and semantic interpretation<br>• Utilized learning mechanisms based on device capabilities, reducing communication costs<br>***Ch:** Rayleigh Fading; **Mt:** Accuracy; **DL:** CNN, FL, Split Learning; **Dt:** MNIST, SST2 [133]* | • Achieved low accuracy on SST2 dataset<br>• Additional overhead for preparation of system |
| | [69] | • Proposed DeepJSCC for wireless image communication using AE-based E2E optimization<br>• Outperforms separate coding schemes, and resistant to channel variation and "cliff effect"<br>***Ch:** AWGN, Rayleigh Fading; **Mt:** PSNR, SSIM; **DL:** AE,CNN; **Dt:** C-10, Kodak* | • Weak against channel mismatch and large image<br>• SNR and CR adaption problem<br>• Baseline (JPEG2000) outperforms at high SNR |
| | [71] | • Proposed DeepJSCC for adaptive-bandwidth image transmission<br>• Developed and validated against multiple description and successive refinement problem<br>***Ch:** AWGN, Rayleigh Fading; **Mt:** PSNR, MS-SSIM; **DL:** AE, CNN; **Dt:** C-10, Kodak, ImageNet [134]* | • Declines when adapting to multiple-channel<br>• High computational complexity and overhead (especially residual transmission scheme)<br>• SNR and CR adaption problem |
| | [72] | • Proposed vision transformer based DeepJSCC scheme<br>• Can adapt to different SNR and bandwidth ratio using single model<br>***Ch:** AWGN; **Mt:** PSNR; **DL:** DL-AE, ViT, Swin transformer; **Dt:** C-10* | • More loss compared to separately trained models<br>• Bandwidth ratio and channel SNR needs to be fed to the model as side information |
| | [73] | • Proposes distributed DeepJSCC with multiple sources for input image<br>• CSI-aware cross attention module to avoid overhead for multiple image reconstruction<br>***Ch:** AWGN, Rayleigh; **Mt:** PSNR. MS-SSIM; **DL:** AE; **Dt:** KITTI12 [135], and KITTI15 [136]* | • Performance drops on high SNR over AWGN<br>• Consider image input of only two sources<br>• SNR and CR adaption problem |
| | [75] | • Proposed DeepJSCC with channel output feedback to improve E2E reconstruction<br>• Reduced delay and overall bandwidth in variable-length transmission<br>***Ch:** AWGN, Rayleigh; **Mt:** PSNR, SSIM, MS-SSIM; **DL:** AE, CNN; **Dt:** Kodak, C-10, ImageNet* | • High computational complexity, inadaptability, suboptimality and non-generalizability<br>• Access to only delayed SNR knowledge based on receiver feedback, and SNR adaption problem |
| Image SemCom | [76] | • Proposed transformer-based feedback-aided DeepJSCC for semantic image transmission<br>• Address inadaptability (over range of SNR), suboptimality and non-generalizability in [75]<br>***Ch:*** *AWGN, Slow fading, Rayleigh, Broadcast;* ***Mt:*** *PSNR, LPIPS;* ***DL:*** *ViT, Siamese, CNN;* ***Dt:*** *CIFAR-10, Kodak, ImageNet, CelebA [137]* | • Assumed passive feedback on received symbols<br>• Relies on encoder for semantic feature extraction<br>• High computational complexity<br>• SNR and CR adaption problem |
| | [78] | • Proposed image SemCom combining ViT and CNN<br>• Introduced novel metrics cossim and Fourier analysis for SemCom<br>***Ch:*** *AWGN, Rayleigh, Over the Air;* ***Mt:*** *PSNR, COSSIM, SSIM, Fourier Analysis;* ***DL:*** *VIT and CNN;* ***Dt:*** *C-10, CIFAR-100 (C-100)[129]* | • Baseline PSNR outperform over for higher bandwidth ratio over AWGN channel<br>• PSNR heavily fluctuates over real channel<br>• High complexity, and encoding/decoding latency |
| | [80] | • Proposed attention DL-based system to overcome SNR adaption problem in DeepJSCC<br>• Solved SNR delay in [75] with attention mechanism for SNR-dependent resource allocating<br>***Ch:*** *AWGN;* ***Mt:*** *PSNR, SSIM;* ***DL:*** *AE, CNN, DNN;* ***Dt:*** *C-10, Kodak, ImageNet* | • Not robust on high-definition images<br>• Complex and real channel models not considered |
| | [82] | • System for high dimensional source transmission by combining NTC and DeepJSCC<br>• Addressed CR-adaption problem using novel adaptive rate transmission<br>• Hyperprior-aided codec refinement technique to enhance E2E transmission performance<br>***Ch:*** *AWGN;* ***Mt:*** *CBR, PSNR, MS-SSIM, and LPIPS;* ***DL:*** *NTSCC,AE, CNN, Transformer ANN;* ***Dt:*** *Kodak, C-10, CLIC2021* | • System performance fluctuates for different dataset (lags behind DeepJSCC on C-10 dataset)<br>• Additional overhead to transmit side information<br>• Complex and real channel models not considered |
| | [83] | • Addressed both SNR and CR adaption problem in DeepJSCC<br>• Achieved variable code rate optimization to meet transmission quality requirements<br>• Achieved data and instance level code optimization for a given PSNR constrain<br>***Ch:*** *AWGN, Slow Rayleigh;* ***Mt:*** *PSNR, Optimized CR, CWM, CWSTD, MSE;* ***DL:*** *AE, CNN, OraNet, DNN;* ***Dt:*** *Kodak24, C-10, C-100* | • Baseline (PDG+LDPC) outperforms across all SNR levels and channels<br>• Coding capability is not better than baseline<br>• Transmission efficiency drops for high target PSNR values |
| | [84] | • Extended DeepJSCC to multipath channels with OFDM blocks<br>• Designed decoder combining DNN and non-trainable (but differentiable) OFDM blocks<br>• Used CNN to directly map source images to complex-valued baseband sample in OFDM<br>***Ch:*** *AWGN, Multipath Fading Channel;* ***Mt:*** *PSNR, SSIM, PAPR, MS-SSIM;* ***DL:*** *AE, GAN, CNN;* ***Dt:*** *Kodak, Open Image Dataset [138]* | • Baseline outperforms (PSNR) for larger images<br>• Few OFDM blocks such as carrier frequency offset, and packet detection are not considered<br>• Signal-clipping during training to reduce PAPR result in performance degradation |
| | [85] | • Proposed channel adaptability for DeepJSCC in OFDM over multipath fading channel<br>• Adaptive to different channel gain and noise power by utilizing CSI and dual-attention<br>***Ch:*** *AWGN, Multipath Fading Channel;* ***Mt:*** *PSNR;* ***DL:*** *AE, CNN;* ***Dt:*** *ImageNet, Kodak* | • Accurate CSI during testing is required to achieve high performance<br>• Perceptual image quality is no evaluated |
| | [87] | • DeepJSCC for transmission over MIMO Rayleigh fading channel with CSI at receiver<br>• First DeepJSCC scheme for multiple antenna system with no CSI available at transmitter<br>• Achieves spatial diversity, and direct mapping of inputs to antenna for spatial multiplexing<br>***Ch:*** *MIMO Rayleigh Fading;* ***Mt:*** *PSNR;* ***DL:*** *AE;* ***Dt:*** *C-10* | • Baseline outperforms for higher number of receiver antenna<br>• Data-driven selection of OSTBCS is not considered in the study |
| | [88] | • ViT-based DeepJSCC for transmission over MIMO channels (open/close-loop MIMO)<br>• Robust against channel estimation errors, channel conditions and antenna numbers<br>***Ch:*** *AWGN, MIMO channels;* ***Mt:*** *PSNR;* ***DL:*** *AE, ViT;* ***Dt:*** *C-10, KODAK, ImageNet, CelebA* | • High computation complexity<br>• SNR adaption problem as separate encoder-decoder pair needs to be trained<br>• CR adaption problem-fixed length JSCC is used |
| | [89] | • Proposed first multi-user DeepJSCC using NOMA for image transmission<br>• Outperformed existing time-division systems<br>***Ch:*** *AWGN, Multiple Access Channel;* ***Mt:*** *PSNR;* ***DL:*** *AE, Siamese, DNN, Curriculum Learning;* ***Dt:*** *C-10* | • Only consider two users for evaluation, and does not consider downlink scenario<br>• Does not validate results with various metrics and datasets |
| | [91] | • Proposed two constellation mapping technique for DeepJSCC system<br>• Outperform traditional QAM constellation mapping<br>***Ch:*** *AWGN;* ***Mt:*** *mIoU, mAP;* ***DL:*** *AE,* ***Dt:*** *Open Image Dataset* | • High computational complexity<br>• Higher order of modulation is not considered<br>• Complex and real channel models not considered |
| | [92] | • DeepJSCC using a finite channel input alphabet and discrete channel input constellation<br>• Learning of constellation for given modulation that outperform convectional constellation<br>***Ch:*** *AWGN, Slow Fading;* ***Mt:*** *PSNR, MS-SSIM;* ***DL:*** *AE, DNN, CNN;* ***Dt:*** *ImageNet, Kodak* | • Inconsistent results – higher order QAM performs better than lower order QAM in high SNR<br>• Baseline outperform for higher order modulation |
| | [93] | • Tradeoff between encoded information and robustness against channel condition<br>• Robust encoding scheme, and task-oriented communication with discrete data-representation<br>***Ch:*** *AWGN;* ***Mt:*** *PSNR, Accuracy, Error Rate, Mutual Information;* ***DL:*** *AE, DNN;* ***Dt:*** *MNIST, C-10* | • Additional overhead by coded redundancy<br>• High memory cost due to the learnable codebook at both transmitter and receiver<br>• Complex and real channel models not considered |
| | [96] | • Proposed DL-based joint transmission – recognition system for IoT devices<br>• Efficient image transmission with balanced bandwidth, reliability, complexity, and accuracy<br>***Ch:*** *AWGN, Rayleigh;* ***Mt:*** *BER, Accuracy;* ***DL:*** *CNN, DNN, ResNet;* ***Dt:*** *C-10* | • Low accuracy at low SNR regions (below 90%)<br>• High computation complexity for offline/online training (not feasible for IoT devices) |
| | [97] | • Image retrieval at network edge, maximizing accuracy under power/bandwidth constraints<br>• Digital and analog schemes, with the analog (DeepJSCC) scheme improving accuracy<br>***Ch:*** *AWGN, Slow Fading Channel;* ***Mt:*** *Accuracy Loss;* ***DL:*** *AE, DNN;* ***Dt:*** *CUHK03 [139], Market-1501 [140], VeRi [141]* | • Computation limitations of IoT devices<br>• Channel model is assumed to be known and remain same during training |
| | [98] | • Task-oriented communication for edge inference from low-end devices to powerful server<br>• Optimized feature extraction, source coding, and channel coding amidst limited bandwidth<br>• The information bottleneck (IB) framework is utilized to formalize a rate-distortion tradeoff<br>***Ch:*** *AWGN;* ***Mt:*** *Error Rate, Latency, PSNR;* ***DL:*** *CNN, ResNet, AE;* ***Dt:*** *MNSIT, C-10, ImageNet* | • High-dimensional data poses a computational bottleneck for IB optimization<br>• E2E training require large data and computational resources, which is not feasible in edge settings<br>• Complex and real channel models not considered |
| | [103] | • Proposed two DeepJSCC (InverseJSCC and GenerativeJSCC) to improve quality of reconstructed images, and address low bandwidth and channel SNR<br>***Ch:*** *AWGN;* ***Mt:*** *LPIPS, MSSIM, PSNR;* ***DL:*** *GAN, AE, CNN;* ***Dt:****ImageNet, CelebA-HQ [142]* | • Used non-differentiable channel<br>• No significant gain in terms of MSSIM and PSNR for high compression ratio |
| Video SemCom | [105] | • Unified video compression and transmission steps using DNN<br>• Optimized bandwidth, and outperformed conventional codecs (H.264+LDPC and H.265+LDPC) in various conditions<br>• ***Ch:*** *AWGN, Fading;* ***Mt:*** *PSNR and MS-SSIM;* ***DL:*** *AE, CNN, RL;* ***Dt:*** *UCF101* [143] | • Low coding efficiency, and achieves lower PSNR value compared to H.265+LDPC<br>• Need for evaluation in practical channels using tools like software-defined radios (SDRs) |
| | [106] | • High-efficiency DeepJSCC for E2E video transmission over wireless channels<br>• Adaptively extracts and transmits semantic features, improving overall performance and efficiency (outperforms H.264/H.265+LDPC and save up to 50% channel bandwidth cost)<br>• High fidelity for both computer vision and human vision tasks<br>***Ch:*** *AWGN, Rayleigh Fading;* ***Mt:*** *PSNR. MS-SSIM, mIoU, Channel Bandwidth;* ***DL:*** *AE, CNN, Swin Transformer;* ***Dt:*** *Vimeo-90k [144], HEVC [145], UVG [146], CamVid [147]* | • High computation as each DVST module requires DNN architecture<br>• Trade-offs between the proposed rate-adaptive mechanism and video quality not discussed |
| | [110] | • Semantic video conference (SVC) network maintain resolution by transmitting keypoints, and improving transmission efficiency with robustness against varying channel condition<br>• SVC integrated a HARQ framework and CSI for enhanced performance and error detection<br>***Ch:****AWGN, Rayleigh Fading;* ***Mt:****Bpp, SSIM, Ploss, AKD;* ***DL:****DNN, CNN;* ***Dt:****VoxCeleb[148]* | • Robustness of SVC with CSI feedback decreases when training and testing environments differ<br>• Loss of detailed expressions due to dependency on keypoint transmission |
| | [112] | • Optimized point cloud video (PCV) streaming on mobile devices using ROI selection, lightweight encoding, and adaptive scheduling<br>• Enhanced real-time PCV streaming performance (improvement of at least 10 FPS to 22 FPS) and overcame resource constraints and dynamic network environments<br>***Ch:*** *3G, 4G,WiFi, and 5G;* ***Mt:*** *Accuracy and Frame Rate;* ***DL:*** *DNN, CNN, RL;* ***Dt:*** *S3DIS [149], 8iVFB [150], Synthetic Dataset* | • Requires significant memory due to caching many neural network models<br>• Primarily focuses on spatial features, temporal features of point cloud frames remain unexplored |
| | [113] | • Proposed novel technique for holographic PCV transmission using E2E network design<br>• The deep reinforcement learning (DRL) based adaptive transmission technique improves latency and reconstruction accuracy while maintaining QoE<br>***Ch:*** *N/A;* ***Mt:*** *FPS, QoE, Time;* ***DL:*** *AE, DRL;* ***Dt:*** *8i Voxelized Point Cloud Dataset* | • Limitations in real-time processing capabilities on mobile devices<br>• High computational and bandwidth requirement for encoding/decoding PCV and model training |
| Audio SemCom | [114] | • DL-based speech transmission system that minimizing semantic level errors<br>• Utilized attention mechanism and squeeze-excitation network, to recover information<br>***Ch:*** *AWGN, Rayleigh, and Rician over Telephone and Multimedia Transmission System;* ***Mt:*** *MSE, SDR, and PESQ;* ***DL:*** *CNN, and SE;* ***Dt:*** *Edinburgh DataShare* | • Overlapping points with conventional system under AWGN channel |
| | [115] | • Proposed DL-based SemCom for speech transmission adaptive to dynamic channel condition<br>• Extracts and transmits semantic features only, improving speech recognition and synthesis<br>***Ch:*** *AWGN, Rayleigh, Rician;* ***Mt:*** *CER, WER, DSD;****DL:*** *CNN, BRNN;* ***Dt:*** *LJSpeech [151]* | • Effects of different speaker information and variations in speech patterns not discussed |
| | [116] | • SemCom for speech and text transmission, transmitting only semantic-relevant information<br>• Their method significantly reduces transmitted symbols, outperforming existing models in accuracy for speech-to-text and quality for speech-to-speech transmissions<br>***Ch:*** *AWGN, and Rayleigh Fading;* ***Mt:*** *WER, Similarity, MOS, MCD-DTW;* ***DL:****CNN, BLSTM, Transformer, GAN;* ***Dt:*** *Librispeech [152], SentencePiece [153]* | • Relies on pre-trained language model<br>• Two-stage training scheme improves training speed, but it affects system adaptability and performance in real-time/dynamic scenarios |
| | [117] | • FL-based AE for audio SemCom, reducing transmission error and overhead<br>• Used wav2vec based AE, to encode-decode audio semantic information, and improve accuracy (Achieves 100 times lower MSE at lower SNR) and noise immunity<br>***Ch:*** *AWGN;* ***Mt:*** *MSE;* ***DL:*** *FL, AE, CNN;* ***Dt:*** *Librispeech* | • Beyond SNR 13dB, LDPC+64QAM achieved better MSE compared to proposed<br>• Does not consider noisy fading channels |
| Multimodal SemCom | [99] | • DL-based multi-user SemCom for transmitting single-modal and multimodal data<br>• Transformer-based frameworks, DeepSC-IR, DeepSC-MT, and DeepSC-VQA, aim to optimize tasks like image retrieval, machine translation, and visual question answering<br>***Ch:*** *MIMO channel, AWGN, Rayleigh Fading, Rician Fading;* ***Mt:*** *Accuracy, BLEU, Recall;* ***DL:*** *Transformer, AE;* ***Dt:*** *Stanford Online Products [154], CUB-200-2011 [155], Cars 196 [156], In-shop Clothes [157], WMT 2018 Chinese-English News [158], CLEVR [159]* | • The system is designed for 3 tasks, adaptability of other types of SemCom tasks or data modalities are not studied<br>• Baseline outperforms in high SNR over AWGN<br>• Only accuracy is evaluated, not quality of data<br>• High Computational Complexity |
| | [119] | • Cross-modal SemCom developed to address polysemy and ambiguity, improving reliability<br>• Multimodal system with video, audio and haptic signal<br>• Development of knowledge-driven quality of experience (QoE) evaluation schemes, considering varying user backgrounds for enhanced SemCom<br>***Ch:*** *N/A;* ***Mt:*** *Mean Absolute Error (MAE);* ***DL:*** *RL, GAN, DNN;* ***Dt:*** *VisTouch* | • Lack of systematic semantic distortion functions affects communication efficiency<br>• Channel/communication environment not given<br>• No knowledge-drive QoE evaluation scheme |
| | [120] | • Cooperative semantic-aware system for IoV, reduce data traffic & alleviate spectrum scarcity<br>• The architecture enhances multiuser communications, emphasizing essential semantics and discarding irrelevant information, improving transmission efficiency and reliability<br>• ***Ch:*** *Single Tap Rayleigh fading;* ***Mt:*** *MSE, Accuracy, mAP;* ***DL:*** *CNN, AE;* ***Dt:*** *VeRi-776* | • Update/management of KB not studied<br>• Only two user is considered for scenario<br>• No mathematical formulation and quantitative analyses of essential semantics extraction |
| | [121] | • Novel MEC structure, utilizing multimodal SemCom for enhanced AR/VR<br>• A CP-DQN method for optimal proactive caching decisions, adapting to user preferences<br>***Ch:*** *Uplink and downlink;* ***Mt:*** *Cache Hit Ratio, Caching Reward, System Cost;* ***DL:*** *DQN, RL* | • High computational complexity<br>• Higher number of user and practical network complexity not discussed |
| | [122] | • Developed a novel generative model for 3D SemCom using SAM and NeRF<br>• The ASCM reduces transmission data and GDCE enhances signal recovery refining CSI<br>***Ch:*** *AWGN, Rician;* ***Mt:*** *NMSE, PSNR, SSIM, Cosine Similarity, BLUE;* ***DL:*** *GAN, NeRF, SAM, ASCM, Diffusion Model (DM);* ***Dt:*** *LLFF, mip-NeRF360 Datasets* | • Multiple advanced models increase overall complexity and computational cost<br>• Reliance on pre-trained models (NeRF and SAM) may limit adaptability to new scenarios |
| | [124] | • SemCom for 3D content transmission, where radiance field models act as semantic knowledge bases to reduce communication overhead and enhance distributed inference<br>• Proposed hierarchical (device-edge-cloud) FL architecture for training radiance field models<br>***Ch:*** *Radio Mapping, Radio Imaging, Environment-aware Communication;* ***Mt:*** *PSNR;* ***DL:*** *NeRF, FL, LSTM;* ***Dt:*** *Local Dataset at End Device, Facial KB, Multi-view Images* | • Training and rendering of radiance field models require significant computational resources<br>• High communication overhead due to frequent exchange of model parameters between participating devices and the central node |

Notation: ***Ch*** are the channel models used, ***Mt*** are evaluation metrics, ***DL*** are DL algorithms, and ***Dt*** are the dataset used in the experiment.

## V. CHALLENGES AND FUTURE DIRECTIONS

Data-driven end-to-end (E2E) semantic communication (SemCom) is revolutionizing numerous industrial sectors and paving the way for advancements beyond 6G networks, impacting areas such as industrial Internet of Things (IIoT), unmanned aerial vehicle (UAV) communications, autonomous vehicles, and smart healthcare systems. These ecosystems can leverage data-driven frameworks for enhanced efficiency and reliability of communication. For instance, in IIoT, employing energy efficient wireless transmission techniques alongside semantic data processing, which emphasizes understanding the meaning of data, can significantly boost both the reliability and efficiency of transmissions. In the context of autonomous vehicles, efficiently sharing perception messages through SemCom-enhanced vehicle-to-everything (V2X) communication can improve network safety, robustness, and efficiency. Similarly, in smart healthcare, by leveraging SemCom for efficient data exchange and energy conservation, AI-driven Internet of Medical Things (IoMT) devices can enhance medical services by analyzing real-time data. Collectively, these advancements pave the way for systems that are smarter, more effective, reliable, and energy efficient. A more detailed discussion on implementing SemCom in various industrial sectors is provided in [13], [14].

While data-driven E2E communication system and SemCom provide a paradigm shift in wireless communication domain, numerous challenges must be overcome to fully implement these systems in practical scenarios. This section discusses the challenges and future direction of deep learning (DL)-based communication systems and their enabling semantic applications.

### A. E2E COMMUNICATION SYSTEM

#### 1) METRICS AND SYSTEM EVALUATION

The dominant performance metrics for DL-based wireless communication systems are block error rate (BLER) and bit error rate (BER), focused on retrieving accurate transmitted signal information. However, in scenarios like SemCom, where not all data transmissions hold equal importance, these metrics may prove inadequate. The crucial factor is maintaining the integrity of semantic information, even amidst errors in data transmission. Consequently, this necessitates a re-evaluation of traditional performance metrics to better align with the requirements of SemCom.

**Potential Research Questions:** How can we quantify the semantic importance of different data packets in a communication stream? What novel metrics can be developed to evaluate the semantic fidelity of transmitted information? How to measure the impact of semantic errors on the overall effectiveness of communication systems?

**Future Direction:** In the evolving landscape of E2E communication systems, future research must prioritize the development of innovative performance metrics tailored for SemCom and E2E scenarios. This involves crafting metrics that accurately reflect the quality of semantic understanding and information fidelity, beyond traditional error rates. Moreover, the implementation of constraint-based training strategies will be crucial for dynamically balancing trade-offs among these new metrics. Such approaches will ensure systems not only maintain efficiency and reliability but also adapt to the nuanced demands of semantic integrity in diverse communication environments. Specific experimental designs may involve simulation-based analysis, where simulation environments model data-driven E2E systems with various levels of information importance for data packets, and both traditional and new metrics are evaluated. Embracing these directions promises to significantly advance the theoretical and practical realms of next-generation wireless communication systems, particularly as we transition into the era of 6G and beyond.

#### 2) DATA LIMITATIONS AND REAL-WORLD DATASET

Wireless communication research frequently relies on simulated data created by researchers themselves. The challenge of obtaining large-scale real-world datasets, aggravated by strict data privacy and protection laws, serves as a significant challenge. Despite the availability of initial datasets for mmWave, SemCom, indoor positioning, etc., standard datasets for a range of application scenarios in wireless communication remain scarce.

**Potential Research Questions:** How do real-world interference and environmental variables affect the performance of various real-world communication systems compared to simulated environments? What are the variations in data transmission efficiency and reliability across different real-world and simulated scenarios?

**Future Direction:** To address the gap between simulation-based methodologies and the need for real-world datasets in E2E systems, future research should focus on developing comprehensive, large-scale datasets that reflect the complexity of real-world scenarios. This endeavor requires a collaborative framework involving academia, industry, and regulatory bodies to overcome data privacy challenges and facilitate data sharing. Specific experimental designs could involve setting up extensive real-world testbeds that capture diverse environmental conditions, user behaviors, and interference patterns, complemented by robust data protection techniques to address privacy concerns. Collaborative research projects could implement secure frameworks such as federated learning (FL), enabling the use of decentralized data from multiple sources without compromising user privacy. By prioritizing the creation of standardized, accessible datasets for diverse applications, researchers can significantly enhance the relevance and applicability of machine learning (ML) model, driving the practical implementation of data-driven technologies across various sectors.

#### 3) SYSTEM COMPLEXITIES IN E2E

The utility of DL-based E2E communication systems, when compared to traditional systems, is still debatable. It is uncertain if these systems can surpass or significantly improve the performance of existing infrastructure in real-world scenarios. As artificial intelligence (AI) and ML become increasingly integrated into wireless communication, the complexity of these systems grows. These complexities, combined with the *black-box* nature of many ML models, complicate the assurance of consistent system performance. The limited computing and storage capacities of mobile devices, along with bandwidth limitations, also restrict the full capabilities of data-driven algorithms. Multi-agent cooperation, which requires various devices to collaborate for optimal performance, introduces further complexity.

**Potential Research Questions:** How can we develop interpretable DL models that provide transparent insights into their decision-making processes? What are the most effective methods for optimizing DL-based E2E systems under constrained wireless resources? How can we design low-complexity, distributed ML algorithms that enable efficient multi-agent cooperation?

**Future Direction:** Considering the complexities identified in E2E systems, future research must delve deeper into the capabilities and optimization of DL-based E2E communication frameworks. It is crucial to develop methodologies that demystify the successes driven by ML in wireless communications, addressing the need for transparent and interpretable ML models. Given the constraints of wireless resources and the challenges in multi-agent communication, pioneering efficient cooperation mechanisms under such limitations becomes essential. Specific experimental designs may involve creating testbeds that simulate real-world wireless environments to evaluate the performance and interpretability of ML models. Additional investigations could focus on the development and assessment of distributed learning frameworks that reduce computational burdens on individual devices while enhancing cooperative communication strategies. Exploring distributed learning and developing low-complexity, ML-based algorithms will be key to managing system complexities and establishing standards, paving the way for more sophisticated, resilient, and efficient E2E communication systems.

#### 4) CROSS-LAYER APPLICATION AND LEARNING

While DL applications have primarily targeted the physical layer (PHY), exploring cross-layer applications introduces potential enhancements. Extending learning across multiple layers can significantly amplify the benefits, bridging gaps between various aspects of the communication system for improved overall performance. While the adaptability of software-based learning has made it a preferred choice, the emergence of hardware learning as a competitive alternative, particularly for handling complex tasks, highlights a pivotal shift. The integration of DL modules tailored with hardware-friendly algorithms into wireless communication systems is increasingly recognized as vital, combining the strengths of both software flexibility and hardware efficiency to meet the demands of advanced communication technologies.

**Potential Research Questions:** How can DL algorithms be optimized for real-time cross-layer adaptation in varying wireless environments? What are the most effective methodologies for integrating hardware-accelerated learning within existing wireless communication frameworks?

**Future Direction:** Future research should prioritize the development of cross-layer applications to unlock comprehensive system benefits, enhancing flexibility and efficiency across communication layers. Simultaneously, the advancement of hardware learning, specifically designed to synergize with wireless communication technologies, becomes essential. Specific experimental designs could involve creating simulation environments that mimics dynamic network conditions to test the performance of cross-layer adaptive algorithms, as well as developing prototypes of hardware-accelerated DL modules and assessing their impact on system latency and throughput in real-world scenarios. Additionally, exploring the integration of SemCom and E2E communication within these cross-layer frameworks could reveal new insights into optimizing information transfer efficiency and accuracy. This dual approach will enable more robust, efficient, and adaptive wireless communication systems, addressing both current challenges and future demands in cross-layer applications.

#### 5) KNOWLEDGE, SECURITY, AND PRIVACY

Integrating prior knowledge into ML architectures, known as knowledge-Driven ML (KDML), streamlines network structures and training processes, thereby improving interpretability. However, effectively incorporating KDML into wireless system design continues to be a challenge. Due to the inherently shared nature of the wireless medium, security and privacy concerns are significantly magnified. Concurrently, the escalating risk of ML models being manipulated or deceived brings forefront the critical need for devising secure ML algorithms that are not only more resilient but also capable of countering these vulnerabilities.

**Potential Research Questions:** How can KDML be tailored to enhance the resilience of ML algorithms against adversarial attacks in wireless networks? What are the most effective methods for embedding prior knowledge into ML models to optimize the performance of E2E communication in varying wireless environments?

**Future Direction:** Advancing research should focus on effectively integrating KDML within wireless systems, addressing the critical need for innovative, secure, and robust ML frameworks. This includes developing novel, interpretable ML algorithms that can navigate the complexities of wireless communication with minimal errors. Furthermore, prioritizing the creation of efficient, hardware-compatible learning models is essential for enhancing system performance under the constraints of limited resources. Security and privacy are paramount for knowledge base (KB) and model protection. Therefore, enhancing resilience of KDML algorithm against adversarial attacks, incorporating several privacy preserving and secure AI techniques like perceptual encryption, differential privacy, secure multiparty computation, and secure communication protocols with robust encryption and authentication mechanisms are essential. Additionally, FL for distributed KDML offers a promising approach to maintaining security and privacy while enabling multi-agent cooperation. Collaborative efforts should also explore secure distributed learning mechanisms to enable seamless multi-agent cooperation, optimizing both the security and functionality of E2E communication networks. Specific experimental designs may explore the implementation of KDML in real-time wireless scenarios, evaluating its impact on error rates, latency, and overall communication efficiency. Developing privacy preserving ML and secure AI algorithms to protect data and model privacy. Additionally, investigations into the use of FL for distributed KDML could provide insights into maintaining security and privacy while enabling multi-agent cooperation.

### B. SEMANTIC APPLICATIONS

#### 1) SEMCOM THEORIES

The absence of a standardized definition, the *block box* nature of DL-based semantic systems, and difficulties in quantifying semantic noise and interference restrict comprehensive performance analysis in SemCom. Furthermore, crafting semantic channel models that account for KB mismatches, goal-driven efficiency, and complex multi-user multimodal communication links continues to be challenging. Establishing unified SemCom theories is crucial for overcoming these obstacles, facilitating autonomous scaling and evolution of SemCom. These theories should encompass semantic-aware networking systems, distributed learning through edge-cloud collaboration, and advanced semantic information processing techniques. Additionally, a comprehensive SemCom theory should incorporate the rate-distortion-perception trade-off, leveraging DL algorithms to achieve accurate semantic representations.

**Potential Research Questions:** How can semantic channel capacities be precisely quantified? What methods can be developed to measure and mitigate semantic noise in real-time? How can semantic information theory be adapted to support multi-user, multimodal communications effectively?

**Future Direction:** Future research within SemCom theory is set to explore crucial areas, including the establishment of a fundamental semantic information theory. This effort involves defining semantic channels and their capacities, along with an in-depth examination of semantic security, efficiency, generalization, and computability [14], [63]. This refined approach aims to lay the groundwork for understanding and enhancing the theoretical underpinnings of SemCom, addressing its core principles and practical implications for future wireless communication systems.

Balancing semantic extraction accuracy with communication overhead poses a significant challenge, especially in environments with many remote users and dynamic communication contexts. Developing systems capable of efficient real-time updates to user KBs without substantial overhead represents a promising research direction. Furthermore, given the reliance of SemCom on transmitting semantically encoded data, ensuring secure wireless communication is paramount. A critical research path involves finding a balance between performance and security, with a particular focus on mitigating the risks posed by adversarial attacks, to enhance both the efficacy and security of SemCom technologies [13]. Specific experimental designs may include creating testbeds for evaluating semantic extraction accuracy in dynamic environments and simulating adversarial attacks to study their impact on SemCom systems. Additionally, implementation of multi-user and multimodal SemCom can further advance the study of its adaptability and efficiency in real work scenarios.

#### 2) SEMCOM ALGORITHMS

SemCom systems face numerous challenges that can restrict their effectiveness. A key issue is the semantic mismatch caused by KB disparities between the sender and receiver, impacting communication clarity. Additionally, deploying DL for SemCom raises issues of interpretability and explainability, with certain DL approaches conflicting with intuitive human understanding. The transmission of semantically encoded messages over unstable wireless channels further risks loss of meaning. This issue becomes more pronounced in real-world conditions that vary significantly from the initial DL training environments, emphasizing the need for SemCom systems, to adaptively maintaining semantic accuracy.

The assessment and design aspects of SemCom also present unique challenges. The current performance metrics for SemCom are insufficient, and there is a pressing need for both objective and subjective unified metrics that can holistically evaluate and compare diverse SemCom techniques [10]. Additionally, the uneven semantic information distribution in SemCom networks demands differential bandwidth allocation. Finding a balance among devices with varying computational and communication capabilities is necessary, since assuming that all devices in these networks possess similar capabilities can lead to inefficiencies [13]. Furthermore, designing these networks to support distributed intelligent systems is crucial for next-generation networks such as 6G, and the primary design challenges stem from the lack of a concrete design blueprint and the complexities involved in sharing users, devices and KBs [14]. Therefore, developing optimal semantic-aware multiple access schemes poses a significant challenge, given the wide variability in devices, traffic patterns, and application requirements.

**Potential Research Questions:** How can adaptive algorithms be developed to minimize semantic mismatches in real-time? What methods can enhance the interpretability and explainability of DL models in SemCom? How can optimal signal detection and interpretation be achieved for multi-user SemCom signals sharing identical resources? How can both explicit and implicit semantic information be integrated to optimize resource distribution in SemCom systems?

**Future Direction:** Multi-user SemCom signals harness the diverse KBs of various SemCom, aiming to save bandwidth in 6G by using shared frequency or time slots [20]. The complexity of optimal signal detection and the interpretation of messages have emerged as crucial fields of study. The use of identical resources for multiple users introduces significant challenges. These include a heightened complexity in decoding semantic information across multiple users and the necessity for receiver KBs to accurately distinguish between user messages. To effectively address these issues, the development and implementation of efficient JSC decoding algorithms are essential. SNR uncertainty plays a pivotal role in the effectiveness of SemCom models, particularly due to variability in noise power (which is a significant factor) [13]. Unlike traditional data caching, which primarily focuses on improving hit rates, semantic caching prioritizes ensuring the precision of the inferred semantic information stored in cache. With the dynamic nature of SemCom, developing innovative algorithms to update and refine semantic caching estimates becomes imperative to adapt to changing contexts. Although most research in SemCom has been centered on explicit semantic information, a significant challenge remains in effectively capturing and utilizing the more nuanced and frequently ignored implicit information. To ensure a balanced distribution of resources, it is essential to develop joint optimization algorithms. These algorithms should adopt a comprehensive approach to SemCom, focusing on integrating both explicit and implicit semantic layers. Specific experimental designs may include creating a simulated environment with varying KB disparities to test adaptive algorithms for minimizing semantic mismatches. Another design could involve comparing different DL approaches using interpretability tools to enhance explainability. Additionally, developing a testing framework with unstable wireless channels can evaluate system resilience. Implementing multi-user scenarios with shared resources can test JSC decoding algorithms. Lastly, implementing and testing joint optimization algorithms in diverse traffic conditions can integrate explicit and implicit semantic information to optimize resource distribution. This holistic strategy is vital for enhancing the overall efficiency and effectiveness of SemCom systems.

### 3) PRACTICAL IMPLEMENTATION

In 6G domain, particularly for Internet of Things (IoT) applications, SemCom stands as a crucial force offering benefits such as lower power and bandwidth usage while maintaining high security [60]. However, incorporating semantic reasoning for error correction introduces latency issues in SemCom transceivers, which challenges the real-time performance required by 6G technologies [14]. Despite the promising capabilities of multimodal and cross-modal SemCom systems, achieving scalability remains a challenge. This is primarily due to the absence of a standardized semantic framework and the computational burdens associated with managing extensive KBs [14], [119], [121].

Furthermore, in a semantic network, the networked KB plays a pivotal role in its intelligence. However, the task of updating such a KB, especially for the diverse communication objects anticipated in 6G and beyond, becomes challenging due to varying computation and storage capabilities. For SemCom to be effectively implemented, it must integrate seamlessly with current communication frameworks. This requires precise performance assessments and the creation of a flexible semantic network that can handle a variety of datasets. However, the effectiveness of DL-based semantic transceivers is limited, as they primarily adapt to static data distributions.

**Potential Research Questions:** How can hardware and physical technologies be enhanced to reduce latency in SemCom transceivers? What novel microelectronic and chip designs can meet the ultra-low latency demands of 6G networks? What systems and techniques can be developed to synchronize and update KBs in real-time across distributed SemCom systems? How can ML algorithms be designed to adapt to dynamic and heterogeneous data sources?

**Future Direction:** To drive SemCom into the forefront of 6G and IoT advancements, future directions must tackle two pivotal challenges. The first challenge involves transcending the limitations of current system-on-chip technologies, which fall short of meeting the ultra-low latency demands critical for 6G networks. This necessitates a leap towards developing cutting-edge microelectronic and chip technologies that can bridge this gap [20]. The second challenge revolves around the dynamic nature of KBs at both the source and destination. Contrary to the prevailing assumption of real-time shared KBs in SemCom research, significant discrepancies still exist. A focused effort on devising methodologies for seamless communication, sharing, and interpretation of semantic information amidst these discrepancies is essential. Specific experimental designs may involve developing prototype chips that integrate advanced materials and architectures, followed by testing their performance in simulated 6G environments. Additional implementation could include creating adaptive synchronization protocols and DL models capable of handling evolving data sets, followed by simulations and real-world deployments to test their robustness and scalability. Such endeavors will not only address the inherent inconsistencies but also pave the way for a more robust and efficient SemCom framework, ensuring successful integration and scalability in future 6G and beyond communication networks and systems.

## VI. CONCLUSION

This study presents a comprehensive review of deep learning (DL)-based end-to-end (E2E) optimization of the physical layer (PHY) in wireless communication systems and the facilitation of semantic applications through E2E learning. It emphasizes the pivotal role of DL in advancing data-driven communication systems, especially with multicarrier waveforms like orthogonal frequency division multiplexing (OFDM), on the path to 6G technologies. DL integration not only enhances PHY layer capabilities for superior communication system performance but also marks a significant shift towards efficient, context-aware semantic applications crucial for intelligent network evolution. Therefore, this paper thoroughly reviews DL strategies for E2E PHY optimization and semantic application enablement across different modalities, evaluating their key contributions and limitations. Despite the promising progress of DL in communication systems, the paper acknowledges the early development stage of these methods and the complexities involved, particularly in practical data-driven wireless communication scenarios. It underlines current challenges and proposes directions for future research, underscoring the imperative for continued DL research in communication systems to achieve fully functional, reliable, secure, bandwidth-efficient, and context-aware networks.

## ACKNOWLEDGMENT

The authors express their sincere appreciation to the editor and anonymous reviewers for their valuable comments and constructive feedback, which significantly improved the quality of this paper.

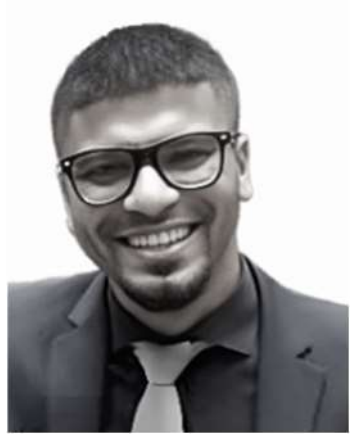

**NAZMUL ISLAM** (Graduate Student Member, IEEE) received the B.S. degree from the Department of Electrical and Computer Engineering, North South University, Dhaka, Bangladesh, in 2017. He is currently pursuing the M.S. degree with the Department of Computer Engineering, Chosun University, Gwangju, Republic of Korea. His research interests include machine learning (ML) in wireless communication, data-driven communication systems, semantic communication, downstream ML applications, cybersecurity, and image processing.

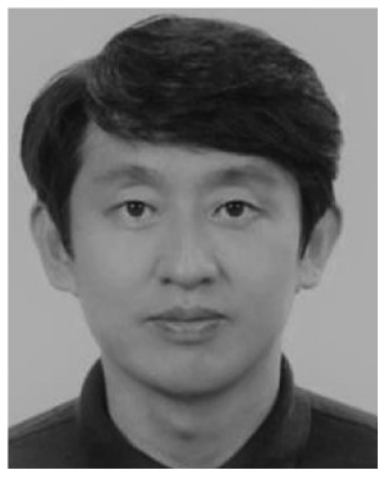

**SEOKJOO SHIN** (Senior Member, IEEE) received the M.S. and Ph.D. degrees from the Department of Information and Communications, Gwangju Institute of Science and Technology (GIST), Republic of Korea, in 1999 and 2002, respectively. He joined the Mobile Telecommunication Research Laboratory, Electronics and Telecommunications Research Institute (ETRI), Republic of Korea, in 2002. In 2003, he joined the Faculty of Engineering, Chosun University, where he is currently a Full Professor with the Department of Computer Engineering. He spent 2009, as a Visiting Researcher with Georgia Tech, USA. His research interests include wireless communication systems and network security and privacy